\documentclass[journal,twoside,final]{IEEEtran}
\usepackage{cite}
\usepackage{graphicx}
\usepackage{amssymb}
\usepackage{amsmath}
\usepackage{amsthm}
\usepackage{cases}
\usepackage{amsfonts}
\usepackage{multirow}
\usepackage{multicol}
\usepackage{mathrsfs}
\usepackage{color}
\usepackage{flushend}
\usepackage{url}
\usepackage{epstopdf}
\usepackage{longtable}
\usepackage{pdflscape}
\allowdisplaybreaks
\usepackage{array}

\usepackage[section]{placeins}

\usepackage{acronym} 

\newtheorem{lemma}{Lemma}
\newtheorem{theorem}{Theorem}

\newtheorem{corollary}{Corollary}

\acrodef{LOS}{line-of-sight}
\acrodef{UE}{user equipment}
\acrodef{BS}{basestation}
\acrodef{RV}{random variable}
\acrodef{i.i.d.}{independent and identically distributed}
\acrodef{i.n.i.d.}{independent but not necessarily identically distributed}
\acrodef{PDF}{probability density function}
\acrodef{CSI}{channel state information}
\acrodef{SNR}{signal-to-noise ratio}
\acrodef{XPD}{cross-polarization discrimination}
\acrodef{XPC}{cross-polarization coupling}
\acrodef{UIU}{unitary-independent-unitary}
\acrodef{SISO}{single-input single-output}
\acrodef{MIMO}{multiple-input multiple-output}
\acrodef{RHS}{right-hand side}
\acrodef{LHS}{left-hand side}
\acrodef{BS}{base station}
\acrodef{3GPP}{3rd Generation Partnership Project}
\acrodef{CSCG}{circularly symmetric complex Gaussian}
\acrodef{DAS}{distributed antenna system}
\acrodef{D-MIMO}{distributed}
\acrodef{C-MIMO}{centralized}
\acrodef{LTE-A}{Long Term Evolution-Advanced}
\acrodef{FDD}{frequency division duplexing}
\acrodef{TDD}{time division duplexing}

\begin{document}

\title{Spectral-Efficiency Analysis of Massive MIMO Systems in Centralized and Distributed Schemes}
\author{Gervais N. Kamga, Minghua~Xia,~\IEEEmembership{Member,~IEEE}, and Sonia~A\"{\i}ssa,~\IEEEmembership{Senior Member,~IEEE}
\thanks{Copyright (c) 2016 IEEE. Personal use of this material is permitted. However, permission to use this material for any other purposes must be obtained from the IEEE by sending a request to pubs-permissions@ieee.org.}
\thanks{
Manuscript received May 31, 2015; revised October 2 and December 30, 2015; accepted January 5, 2016. This work was supported by a Discovery Grant from the Natural Sciences and Engineering Research Council (NSERC) of Canada. The associate editor coordinating the review of this manuscript and approving it for publication was Dr. V. Raghavan.}
\thanks{
G. N. Kamga and S.~A\"{\i}ssa are with the Institut National de la Recherche Scientifique (INRS), University of Quebec, Montreal, QC, H5A 1K6, Canada (e-mail:  \{kamga, aissa\}@emt.inrs.ca).}
\thanks{
M. Xia is with the School of Electronics and Information Engineering, Sun Yat-sen University (East Campus), Guangzhou, 510006, China (e-mail: xiamingh@mail.sysu.edu.cn). He was with the Institut National de la Recherche Scientifique (INRS), University of Quebec, Montreal, QC, H5A 1K6, Canada.}
\thanks{
Color versions of one or more of the figures in this paper are available online at http://ieeexplore.ieee.org.}
\thanks{
Digital Object Identifier XXX}
}

\markboth{IEEE Transactions on Communications, accepted for publication, January 2016.} {Kamga \MakeLowercase{\textit{et al.}}: Spectral-Efficiency Analysis of Massive MIMO Systems in Centralized and Distributed Schemes}

\maketitle

\pubid{XXXX-XXXX}

\pubidadjcol

\begin{abstract}
\noindent This paper analyzes the spectral efficiency of massive \ac{MIMO} systems in both centralized and distributed configurations, referred to as C-MIMO and D-MIMO, respectively. By accounting for real environmental parameters and antenna characteristics, namely, path loss, shadowing effect, multi-path fading and antenna correlation, a novel comprehensive channel model is first proposed in closed-form, which is applicable to both types of MIMO schemes. Then, based on the proposed model, the asymptotic behavior of the spectral efficiency of the \ac{MIMO} channel under both the centralized and distributed configurations is analyzed and compared in exact forms, by exploiting the theory of very long random vectors. Afterwards, a case study is performed by applying the obtained results into MIMO networks with circular coverage. In such a case, it is attested that for the D-MIMO of cell radius $r_{\mathrm{c}}$ and circular antenna array of radius~$r_{\mathrm{a}}$, the optimal value of~$r_{\mathrm{a}}$ that maximizes the average spectral efficiency is accurately established by $r_{\mathrm{a}}^{\mathrm{opt}}=r_{\mathrm{c}}/1.31$. Monte Carlo simulation results corroborate the developed spectral-efficiency analysis.
\end{abstract}


\acresetall 

\section{Introduction}
\label{sec:intro}

\begin{figure}[tb]
\centering
\includegraphics[width=3.5in]{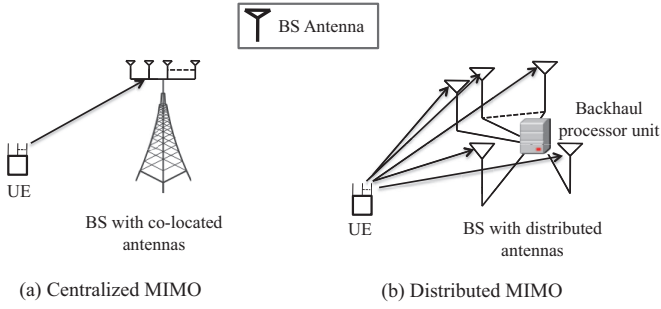}
\caption{Two configurations of point-to-point massive \ac{MIMO} systems: (a) \ac{C-MIMO}, where antennas are co-located at the BS and the UE sides and, thus, distances from UE to BS antennas are almost identical; (b) \ac{D-MIMO}, where BS antennas are deployed at different geographical locations while connected together through the backhaul processor unit, implying that the distances from UE to BS antennas are different.}
\label{fig:system_model_C_MIMO_D_MIMO}
\end{figure}

\IEEEPARstart{M}{assive} \ac{MIMO} communication technique, where tens or a few hundred antennas are deployed at either or both ends of a wireless link,  promises significant performance gains in terms of spectral efficiency, energy efficiency, security and reliability compared with conventional \ac{MIMO} \cite{Larsson2014}, and is becoming a cornerstone of future 5G systems \cite{AndrewsJSAC1406}. To implement massive \ac{MIMO} in wireless networks, two different schemes can be adopted (see, e.g., \cite{Bjornson2015, Firouzabadi2011, Dongming2013}): \ac{C-MIMO}, where antennas are co-located at both the transmit (Tx) and the receive (Rx) sides as illustrated in Fig.~\ref{fig:system_model_C_MIMO_D_MIMO}-a (which is essentially equivalent to conventional \ac{MIMO} system), and \ac{D-MIMO}, where \ac{BS} antennas are deployed at different geographical locations while connected together through high-capacity backhaul links such as fibre-optic cables, as shown in Fig.~\ref{fig:system_model_C_MIMO_D_MIMO}-b.

From a practical point of view, \ac{C-MIMO} is more easy to mathematically analyze and physically deploy, compared with \ac{D-MIMO}. In fact, unlike the former, the latter suffers from different degrees of path losses caused by different access distances to different distributed antennas, which makes the performance analysis and design more challenging. Also, since the location of antennas in \ac{D-MIMO} has a significant effect on the system performance, optimization of the antenna locations is crucial \cite{Xinzheng2009,Firouzabadi2011}. This task may become very challenging because of the large numbers (massive) of Tx/Rx antennas. On the other hand, in practice, arbitrary antenna locations or optimal topology may lead to a prohibitive cost for the backhaul component, as well as installation cost for the distributed setting.

\ac{D-MIMO} technique, however, exhibits several advantages compared with \ac{C-MIMO}, such as lower transmit power, higher multiplexing gain, higher spectral efficiency, enhanced coverage area and ease of network planning \cite{XiaJSAC1996,Schuh2002}. As such, both \ac{C-MIMO} and \ac{D-MIMO} represent promising choices for practical implementation of massive \ac{MIMO} technique, each depending on potentially preferable criteria mentioned above.

No matter whether the centralized or distributed configuration is concerned, to capture the propagation characteristics and to understand the system performance and behaviour in real physical environments, two fundamental tasks are to {\it i)} develop an analytical channel model, where path loss, shadowing effect and multi-path fading are accounted for; and to {\it ii)} conduct analytical performance evaluation and assess key factors that determine system performance. In particular, for \ac{D-MIMO} systems, different path losses and shadowing effects w.r.t. different \ac{BS} antennas are critical to the realization of Tx/Rx diversity. In addition, antenna correlation is inherent to the realization of massive \ac{MIMO},  because of the lack of sufficient physical space to separate the large number of antennas in case they are co-located.

\pubidadjcol

In practice, the performance of point-to-point massive \ac{MIMO} serves as a benchmark for further performance evaluation in multi-user settings. Also, point-to-point massive \ac{MIMO} finds wide applications, e.g., high-speed wireless backhaul link between \acp{BS} \cite{Chockalingam2014}. However, despite the extreme importance of point-to-point massive \ac{MIMO}, there is no existing work that successfully accounts for all the aforementioned parameters (i.e., path loss, shadowing effect, multi-path fading and antenna correlation), while developing channel model and conducting closed-form performance analysis. In particular, it was shown in \cite{Chuah_TIT_2002} that for \ac{MIMO} channels, the spectral efficiency grows linearly with the minimum between the numbers of Tx and Rx antennas, even if they tend to infinity. The asymptotic result when the number of antennas at only one side goes to infinity was reported in \cite{Telatar1999, Goldsmith03}. In \cite{Chuah_TIT_2002,Telatar1999, Goldsmith03}, only basic multi-path fading was considered whereas path loss, shadowing and antenna correlation were ignored. In \cite{TulinoJ_IT2005,Veeravalli2005,Raghavan2011}, the capacity of correlated multi-antenna channels was studied, in both regimes of finite numbers of antennas (in \cite{TulinoJ_IT2005}) and large numbers of antennas (in \cite{TulinoJ_IT2005,Veeravalli2005,Raghavan2011}), where only the Rayleigh fading and the antenna correlation were considered. Recently, in \cite{KamgaTWCOM2014} and its companion conference version \cite{Kamga_ICC2015}, a comprehensive channel model consisting of path loss, shadowing, multi-path fading, antenna correlation and polarization was firstly developed. Then, an upper bound on the ergodic capacity of point-to-point \ac{C-MIMO} was derived, by using the Hadamard's determinant inequality, and further asymptotically analyzed in the sense of larger number of Tx and/or Rx antennas.

Compared to our previous work \cite{KamgaTWCOM2014, Kamga_ICC2015}, the major contributions of this paper are summarized as follows.

1) This paper develops a general channel model suitable for both massive \ac{C-MIMO} and \ac{D-MIMO}, where major environmental parameters and antenna physical parameters are accounted for. As apposed to the Kronecker correlation model used in \cite{Kamga_ICC2015}, this paper uses the Weichselberger model which is more accurate than the former (as detailed later in Section \ref{sec:channel_model}). Moreover, although various parameters needed in channel modeling have been partially considered to some extent in the open literature, key channel parameters, namely, path loss, shadowing, multi-path fading and antenna correlation, are concurrently taken into account and their effects on spectral efficiency are investigated in this paper.

2) The asymptotic behavior of the spectral efficiency of \ac{C-MIMO} and \ac{D-MIMO} is analyzed, in the sense of large number of Rx antennas. More specifically:
\begin{enumerate}
\item[a)] We first extend the law of large numbers for very long random vectors with \ac{i.n.i.d.} entries, to the more general case of very long random vectors with {\it weighted} \ac{i.n.i.d.} entries, where a condition is imposed onto the weights to guarantee the convergence in probability.

\item[b)] Then, two target matrices are introduced in the expressions of the spectral efficiency: $\mathbf{M}_\mathrm{C}$ for \ac{C-MIMO} and $\mathbf{M}_\mathrm{D}$ for \ac{D-MIMO}. Afterwards, the above results on the law of large numbers for very long random vectors with {\it weighted} \ac{i.n.i.d.} entries are exploited to derive the asymptotic expressions of the entries of $\mathbf{M}_\mathrm{C}$ and of $\mathbf{M}_\mathrm{D}$, w.r.t. the number of Rx antennas $N_R$ where $N_R \to \infty$.

\item[c)] The resulting asymptotic behavior is then applied to derive the intended spectral efficiency, yielding novel expressions from which new insights into the system performance can be gained. In particular, our results show that, {\it i)} \ac{D-MIMO}  does not always outperform \ac{C-MIMO}  in terms of spectral efficiency; {\it ii)} \ac{D-MIMO} exhibits a higher multiplexing gain than that of \ac{C-MIMO}, up to $N_T \times N_R$ where $N_T$ denotes the number of Tx antennas; and {\it iii)} the performance of massive \ac{MIMO} on the uplink is mainly determined by the correlation characteristics at the Tx side instead of the Rx side, given that the Weichselberger correlation model is applied.
\end{enumerate}

3) A case study is performed by applying the obtained results in a pertinent scenario where \ac{D-MIMO} adopts a circular topology, and several key insights into the system performance and optimal antenna deployment are gained. In particular, it is demonstrated that for \ac{D-MIMO} with circular topology of cell radius $r_{\mathrm{c}}$ and circular antenna array of radius $r_{\mathrm{a}}$, the optimal value of $r_{\mathrm{a}}$ that maximizes the average spectral efficiency is given by $r_{\mathrm{a}}^{\mathrm{opt}}=r_{\mathrm{c}}/1.31$.

To detail the aforementioned contributions, the following content of the paper is organized as follows. Section \ref{Sec:modeling} develops the channel model suitable for both \ac{C-MIMO} and \ac{D-MIMO}, and Section \ref{sec:asymptotic_capacity} derives the associated asymptotic spectral efficiency. Afterwards, the case study is conducted in Section \ref{sec:location_optimization}. Section \ref{sec:simulations} presents numerical results pertaining to the developed analyses, in comparison with Monte Carlo simulation results. Section \ref{sec:conclusions} concludes the paper and, finally, some detailed derivations are relegated to appendices.

\textit{Notation:}
In the paper, scalars are represented by lowercase letters like $h$, whereas vectors and matrices are represented by bold lowercase and uppercase letters, like $\mathbf{h}$ and $\mathbf{H}$, respectively. The row vector $\mathbf{h}$ with size $n$ is written as $\mathbf{h} = [h_1, h_2, \cdots, h_n]$ and the $(i, j)^{\mathrm{th}}$ entry of $\mathbf{H}$ is denoted by $[\mathbf{H}]_{i, j}$. The subscript $n\times m$ in $\mathbf{A}_{n\times m}$ means the size of $\mathbf{A}$, i.e., $\mathbf{A} \in \mathcal{C}^{n \times m}$. $\mathbf{1}_n$ refers to the $n \times 1$ column vector with all entries being unity, and $\mathbf{I}_n$ denotes the identity matrix of size $n \times n$. Operators $(.)^{\rm T}$, $(.)^{\rm H}$, $\det(.)$ and $\odot$ refer to the transpose, Hermitian transpose, determinant and Hadamard product, respectively. $\mathrm{Pr}(.)$, $\mathbb{E}\{.\}$ and $\mathrm{Var}\{.\}$ stand for mathematical probability, expectation and variance, respectively.
\section{Massive MIMO System and Channel Modeling}
\label{Sec:modeling}
\subsection{System Models of \ac{C-MIMO} and \ac{D-MIMO} }
\label{sec:D-MIMO_and_C-MIMO}
We consider the uplink of a multi-user massive \ac{MIMO} system, where the \ac{BS} is equipped with $N_R$ Rx antennas while each \ac{UE} is equipped with $N_T$ Tx antennas. In the centralized setting as illustrated in Fig.~\ref{fig:system_model_C_MIMO_D_MIMO}-a, the $N_R$ antennas of the \ac{BS} are co-located, whereas in the distributed scheme shown in Fig.~\ref{fig:system_model_C_MIMO_D_MIMO}-b, the $N_R$ antennas of the \ac{BS} are separately distributed in geography. In both settings, it is assumed that the number of antennas at the \ac{BS} is large, i.e., the value of $N_R$ is on the order of tens or even hundreds, and that $N_R > N_T$ since the number of Tx antennas at \ac{UE} is usually not large due to the physical size limitations. Since this paper focuses on a comprehensive channel model used for the system performance analysis, it is assumed that there is no hardware imperfections at the \ac{BS} antenna array.\footnote{For the reader interested in hardware imperfections of antenna array,  please refer to, e.g., \cite{Bjornson2015}.}  Further, it is assumed that in the \ac{D-MIMO} scheme, high-capacity backhaul links such as fibre-optic cables connect the \ac{BS} antennas, which cooperate perfectly with each other \cite{Firouzabadi2011,Dongming2013}.
\subsection{Channel Models}
\label{sec:channel_model}
In both \ac{C-MIMO} and \ac{D-MIMO} systems, the physical channel between both ends of any communication link is assumed to be subject to path loss, shadowing and multi-path fading. In the centralized scheme, the path-loss components on the radio link between the \ac{UE} antennas and the \ac{BS} antennas are \ac{i.i.d.}, since in this setting, antennas at either end (transmitter or receiver) are at the same location. In the distributed scheme, on the other hand, since different \ac{BS} antennas are deployed at different geographical locations, the path-loss components of radio links between the \ac{UE} and the antennas of the \ac{BS} are \ac{i.n.i.d.}. With the system model described above, a channel model applicable to both \ac{C-MIMO} and \ac{D-MIMO} systems, $\mathbf{H} \in \mathcal{C}^{N_R \times N_T}$, can be explicitly given by
\begin{equation}
\label{eq:channel_model_pres}
\mathbf{H} = \mathbf{D}^{\frac{1}{2}}\mathbf{H}_{0},
\end{equation}
where $\mathbf{H}_{0} \in \mathcal{C}^{N_R \times N_T}$ models multi-path fading while $\mathbf{D}\in \mathcal{R}^{N_R \times N_R}$ represents path loss and shadowing effect. With the observations right before \eqref{eq:channel_model_pres}, it is readily shown that $\mathbf{D}$ in \eqref{eq:channel_model_pres} is a diagonal matrix given by
\begin{numcases}{\mathbf{D}=}
\label{eq:entries_of_C_large-scale_matrix}
d^{-\nu} \varphi \, \mathbf{I}_{N_R},
& \mbox{\ac{C-MIMO}};
 \\
\label{eq:entries_of_D_large-scale_matrix}
\mathrm{diag}\left\{d_i^{-\nu}{\varphi}_i\right\}_{i=1}^{N_R},
& \mbox{\ac{D-MIMO}},
\end{numcases}
where in \eqref{eq:entries_of_C_large-scale_matrix}, $d$ and $\varphi$ denote the Euclidean distance and the shadowing effect pertaining to the link between the \ac{UE} and the \ac{BS} of a C-MIMO system, respectively; and where in \eqref{eq:entries_of_D_large-scale_matrix}, $d_i$ and $\varphi_i$ refer to the Euclidean distance and the shadowing effect pertaining to the link between the \ac{UE} and the $i^{\mathrm{th}}$ \ac{BS} antenna of a D-MIMO system, for all $i \in \{1, 2,\ldots,N_R\}$. In \eqref{eq:entries_of_C_large-scale_matrix} and \eqref{eq:entries_of_D_large-scale_matrix}, $\nu > 2$ is the path-loss exponent. By using a similar methodology as detailed in \cite[Sec. II.A]{Kamga_ICC2015}, $\varphi$ shown in \eqref{eq:entries_of_C_large-scale_matrix} can be well described by a Gamma distribution. Accordingly, the \ac{PDF} of $\varphi$ can be written as
\begin{equation}
\label{eq:gamma_pdf_shadowing_C_MIMO}
f_{\varphi}(x)=\frac{1}{\Gamma(\alpha)}\left(\frac{\alpha}{\Omega}\right)^{\alpha}x^{\alpha-1}\exp{\left(-\frac{\alpha}{\Omega}x\right)},
\end{equation}
where $\Gamma(x) = \int_0^x{{t^{x-1}e^{-t}}\,\mathrm{d}t}$ denotes the Gamma function, $\alpha > 0.5$ inversely reflects the shadowing severity and $\Omega$ is the average power of the shadowing effect. For the sake of brevity, the Gamma distribution in the form of \eqref{eq:gamma_pdf_shadowing_C_MIMO} is shortly denoted $\mathcal{G}(\alpha,\Omega/\alpha)$, with $\alpha$ and $\Omega/\alpha$ being the shape parameter and the scaling factor, respectively. Accordingly, $\varphi_i$ shown in \eqref{eq:entries_of_D_large-scale_matrix} is distributed according to
\begin{equation}
\label{eq:gamma_pdf_shadowing_D_MIMO}
f_{\varphi_i}(y)=\mathcal{G}\left(\alpha_i, \frac{\Omega}{\alpha_i}\right),				
\end{equation}
where $\alpha_i$ denotes the shadowing parameter pertaining to the link between the \ac{UE} and the $i^{\mathrm{th}}$ \ac{BS} antenna, for all $i \in \{1, 2,\ldots,N_R\}$.

If antenna correlation at the Tx and Rx sides is considered and modelled by the well-known Kronecker model, namely,
\begin{equation}
\label{eq:Kronecker_model}
\mathbf{H}_0 \triangleq \mathbf{\Theta}_{R}^{\frac{1}{2}}\,\widehat{\mathbf{H}}\left(\mathbf{\Theta}_{T}^{\frac{1}{2}}\right)^{\rm H},
\end{equation}
where $\mathbf{\Theta}_{T} \in \mathcal{C}^{N_T \times N_T}$ and $\mathbf{\Theta}_{R} \in \mathcal{C}^{N_R \times N_R}$ refer to the correlation matrices at the transmitter and the receiver, respectively; and where the $(i, j)^\mathrm{th}$ entry of matrix $\widehat{\mathbf{H}}$, for all $i \in \{1, 2,\ldots,N_R\}$ and $j \in \{1, 2,\ldots,N_T\}$, follows a \ac{CSCG} distribution:
\begin{equation}
\label{eq:entries_of_small-scale_matrix}
 [\widehat{\mathbf{H}}]_{i, j} \sim \mathcal{CN}(0,1),
\end{equation}
then, substituting \eqref{eq:entries_of_C_large-scale_matrix}, \eqref{eq:entries_of_D_large-scale_matrix} and \eqref{eq:Kronecker_model} into \eqref{eq:channel_model_pres} yields the channel model
\begin{numcases}{\mathbf{H}=}
\label{eq:global_channel_C-MIMO}
d^{-\frac{\nu}{2}}{\varphi}^{\frac{1}{2}}\,\mathbf{\Theta}_{R}^{\frac{1}{2}}\,\widehat{\mathbf{H}}\left(\mathbf{\Theta}_{T}^{\frac{1}{2}}\right)^{\rm H}, & \mbox{\ac{C-MIMO}};
\\
\label{eq:global_channel_D-MIMO}
\mathrm{diag}\left\{d_i^{-\frac{\nu}{2}}{\varphi}_i^{\frac{1}{2}}\right\}_{i=1}^{N_R} \,\widehat{\mathbf{H}}\,\left(\mathbf{\Theta}_{T}^{\frac{1}{2}}\right)^{\rm H},
 & \mbox{\ac{D-MIMO}}.
\end{numcases}
It is noted that $\mathbf{\Theta}_{R}$ in \eqref{eq:global_channel_C-MIMO} denotes the correlation matrix at the Rx antennas in the C-MIMO scheme. As far as D-MIMO is concerned, however, Rx antennas at BS are well separated in geography and, thus, correlation between them is negligible. Accordingly,  $\mathbf{\Theta}_{R}$ in \eqref{eq:global_channel_C-MIMO} reduces to an identity matrix in the \ac{D-MIMO} scheme, as implied by \eqref{eq:global_channel_D-MIMO}.

By recalling the exponential correlation model widely used between antenna elements \cite{Oestges2008, Maaref2008}, the entries of $\mathbf{\Theta}_R$ and $\mathbf{\Theta}_{T}$ in the model above can be explicitly given by
\begin{equation}
\label{eq:corr_matrix}
[\mathbf{\Theta}_I]_{k, l} = {\theta}_{I}^{|k-l|},~\forall k, l \in  \left\{1,2,\cdots, N_I\right\},
\end{equation}
where $I \in \left\{T, R \right\}$ and ${\theta}_{I}=e^{-L_I/\Delta_I}$, with $L_I$ being the sub-array spacing at the \ac{UE} if $I=T$ and at the \ac{BS} if $I=R$, and $\Delta_I$ denoting characteristic distances proportional to the spatial coherence distance at each side \cite{Oestges2008}.

Though the Kronecker model shown in \eqref{eq:Kronecker_model} is by far the most popular correlation model used in conventional \ac{MIMO} systems, mainly due to its simplicity and analytical tractability \cite{Costa2010}, the accuracy of this model suffers from some limitations (cf. \cite{WeichselbergerTWC2006, Xiang2014, Ozcelik2003, Raghavan2010} and references therein), especially in massive MIMO systems. For instance, it may underestimate the channel capacity. As an alternative to the Kronecker model, and inspired by the latter, the Weichselberger model was proposed \cite{Weichselberger_Thesis, WeichselbergerTWC2006},\footnote{It is noteworthy that the ``Weichselberger model'' is called by different names in the MIMO literature, e.g., non-separable correlation model, UIU model, virtual representation, etc. For more details, see, e.g., \cite{TulinoJ_IT2005, Veeravalli2005, Raghavan2010,Raghavan2011}.} which is a reformulation of the Kronecker model and has been shown to be more accurate. The Weichselberger correlation model is derived as follows. By first applying the eigenvalue decomposition in matrix theory to $\mathbf{\Theta}_{T}$ and $\mathbf{\Theta}_{R}$ shown in \eqref{eq:Kronecker_model}, similarly as in \cite[Eq. (12)]{Kamga_ICC2015}, while recalling that $\mathbf{\Theta}_{R}$ reduces to the identity matrix $\mathbf{I}_{N_R}$ in the \ac{D-MIMO} scheme as discussed right after \eqref{eq:global_channel_D-MIMO}; then by exploiting the Weichselberger reformulation of the Kronecker model shown in \eqref{eq:Kronecker_model}, we obtain the so-called \ac{UIU} formulation \cite[Ch. 6.4.3]{Weichselberger_Thesis}. Accordingly, \eqref{eq:Kronecker_model} can be rewritten as
\begin{numcases}{\mathbf{H}_0 =}
\label{eq:UIU_formulation_C-MIMO}
 \mathbf{U}_{R}^{\frac{1}{2}}(\mathbf{G}_{\mathrm{C}}\odot \widehat{\mathbf{H}})\left(\mathbf{U}_{T}^{\frac{1}{2}}\right)^{\rm H},
& \mbox{\ac{C-MIMO}};
\\
\label{eq:UIU_formulation_D-MIMO}
(\mathbf{G}_{\mathrm{D}}\odot \widehat{\mathbf{H}})\left(\mathbf{U}_{T}^{\frac{1}{2}}\right)^{\rm H},
 & \mbox{\ac{D-MIMO}},
\end{numcases}
where $\mathbf{U}_{T} \in \mathcal{C}^{N_T \times N_T}$ and $\mathbf{U}_{R} \in \mathcal{C}^{N_R \times N_R}$ are unitary matrices, and where $\mathbf{G}_{\mathrm{C}}$ and $\mathbf{G}_{\mathrm{D}}$ are defined as
\begin{equation}
\label{eq:eigenmode_coupling_elt_wise_square-root}
\mathbf{G}_{\mathrm{J}}=\boldsymbol{\lambda}_{R}^{\frac{1}{2}}\left(\boldsymbol{\lambda}_{T}^{\frac{1}{2}}\right)^{\rm T},~\forall \mathrm{J} \in \{\mathrm{C}, \mathrm{D}\}
\end{equation}
with the vectors
\begin{equation}
\label{eq:singular_value_vectors_Tx}
\boldsymbol{\lambda}_{T}\triangleq\begin{bmatrix} \lambda_{T,1} & \lambda_{T,2} & \ldots & \lambda_{T,N_T} \end{bmatrix}^{\rm T},
\end{equation}
\begin{equation}
\boldsymbol{\lambda}_{R}\triangleq\begin{bmatrix}\lambda_{R,1} & \lambda_{R,2} & \ldots & \lambda_{R,N_R}\end{bmatrix}^{\rm T},
\label{eq:singular_value_vectors_Rx}
\end{equation}
consisting of the eigenvalues of the matrices $\mathbf{\Theta}_{T}$ and $\mathbf{\Theta}_{R}$, respectively.

Finally, by virtue of the new expression of $\mathbf{H}_0$ shown in \eqref{eq:UIU_formulation_C-MIMO} and \eqref{eq:UIU_formulation_D-MIMO}, Eqs. \eqref{eq:global_channel_C-MIMO} and \eqref{eq:global_channel_D-MIMO} can be rewritten, yielding a general and unified channel model suitable for \ac{C-MIMO} and \ac{D-MIMO} systems given by
\begin{numcases}{\mathbf{H}\!\!=\!\!}
\label{eq:global_channel_C-MIMO-final}
\!\!d^{-\frac{\nu}{2}}{\varphi}^{\frac{1}{2}}\,\mathbf{U}_{R}^{\frac{1}{2}}(\mathbf{G}_{\mathrm{C}}\odot \widehat{\mathbf{H}})\left(\mathbf{U}_{T}^{\frac{1}{2}}\right)^{\rm H},\,~~~~\mbox{\ac{C-MIMO}};
\\
\label{eq:global_channel_D-MIMO-final}
\!\!\!\mathrm{diag}\!\left\{\!d_i^{-\frac{\nu}{2}}{\varphi}_i^{\frac{1}{2}}\right\}_{i=1}^{N_R} \!(\mathbf{G}\odot \widehat{\mathbf{H}})\!\left(\!\mathbf{U}_{T}^{\frac{1}{2}}\right)^{\rm H},
 \mbox{\ac{D-MIMO}}.
\end{numcases}
As mentioned in the Introduction, for ease of mathematical tractability, various analytical models were developed by accounting for only partial key parameters needed in MIMO channel modeling in the open literature, whereas major channel parameters including path loss, shadowing, multi-path fading and antenna correlation are all involved in  \eqref{eq:global_channel_C-MIMO-final}--\eqref{eq:global_channel_D-MIMO-final}.

\section{Spectral Efficiency Analysis}
\label{sec:asymptotic_capacity}

In this section, we derive the asymptotic spectral efficiency of \ac{C-MIMO} and \ac{D-MIMO} mathematically described by \eqref{eq:global_channel_C-MIMO-final} and \eqref{eq:global_channel_D-MIMO-final}, respectively.\footnote{It is noteworthy that the expression ``spectral efficiency'' is, in general, a rather abused term in the \ac{MIMO} literature. In the context of low-SNR analysis, spectral efficiency stands for capacity-per-unit-energy type arguments, while in the context of high-SNR analysis, it stands for the rate. In this paper, ``spectral efficiency'' is used in the context of rate/throughput.} To this end, two target matrices are first introduced in the expression of the spectral efficiency: $\mathbf{M}_\mathrm{C}$ for \ac{C-MIMO} and $\mathbf{M}_\mathrm{D}$ for \ac{D-MIMO}. Then, the asymptotic expressions of the entries of $\mathbf{M}_\mathrm{C}$ and of $\mathbf{M}_\mathrm{D}$ are explicitly derived, w.r.t. $N_R$ where $N_R \to \infty$. Afterwards, the resulting asymptotic behavior is applied to derive the intended spectral efficiency, which yields novel expressions from where several new insights into the system performance are gained.

\subsection{Instantaneous Spectral Efficiency of MIMO Channel}
\label{subsec:channel_capacity}

It is assumed that the \ac{BS} perfectly knows the \ac{CSI} whereas no CSI is available at the \ac{UE}.\footnote{In the context of massive MIMO systems, channel reciprocity is widely exploited to estimate the channel response on the uplink and then use the acquired CSI for both uplink Rx combining and downlink Tx precoding of payload data, provided that the system operates in TDD mode \cite{Bjornson15}.  If the system operates in FDD mode, the uplink and downlink use different frequency bands and channel reciprocity cannot be harnessed. In such a case, the CSI acquisition becomes much more challenging. For more information, the interested reader is referred to the survey \cite{Bjornson15}.} Accordingly, the total Tx power, $P$, is uniformly allocated across the Tx antennas of the \ac{UE}. Then, by recalling that $N_T < N_R$, the instantaneous spectral efficiency of the \ac{MIMO} channel in the unit of bit/s/Hz is readily given by
\begin{equation}
 C = \log_2 \left[\det\left(\mathbf{I}_{N_T}+\frac{\rho}{N_T}{\mathbf{H}}^{\rm H}\mathbf{H}\right)\right],
\label{eq:capacity_definition}
\end{equation}
where $\rho$ denotes the average Tx \ac{SNR}.

Let $C_\mathrm{C}$ and $C_\mathrm{D}$ denote the spectral efficiency pertaining to \ac{C-MIMO} and \ac{D-MIMO}, respectively. By substituting \eqref{eq:global_channel_C-MIMO-final} into \eqref{eq:capacity_definition} for \ac{C-MIMO} while \eqref{eq:global_channel_D-MIMO-final} into \eqref{eq:capacity_definition} for \ac{D-MIMO}, and using the identity $\det(\mathbf{I}_{n}+\mathbf{A}_{n\times m}\mathbf{B}_{m\times n})=\det(\mathbf{I}_{m}+\mathbf{B}_{m\times n}\mathbf{A}_{n\times m})$, after performing some algebraic manipulations we obtain
\begin{equation}
C_\mathrm{C}= \log_2\left[\det\left(\mathbf{I}_{N_T}+\frac{\rho}{N_T}\mathbf{M}_\mathrm{C}\right)\right],
\label{eq:capacity_definition_C-MIMO-1}
\end{equation}
\begin{equation}
C_\mathrm{D}
= \log_2\left[\det\left(\mathbf{I}_{N_T}+\frac{\rho}{N_T}\mathbf{M}_\mathrm{D}\right)\right],
\label{eq:capacity_definition_D-MIMO-1}
\end{equation}
where
\begin{equation} \label{eq:M_C}
\mathbf{M}_\mathrm{C}
\triangleq d^{-\nu}\,{\varphi}\left(\mathbf{G}_{\mathrm{C}}\odot \widehat{\mathbf{H}}\right)^{\rm H}\left(\mathbf{G}_{\mathrm{C}}\odot \widehat{\mathbf{H}}\right),
\end{equation}
\begin{equation} \label{eq:M_D}
\mathbf{M}_\mathrm{D}\triangleq \left(\mathbf{G}_{\mathrm{D}}\odot \widehat{\mathbf{H}}\right)^{\rm H}\mathbf{D}\left(\mathbf{G}_{\mathrm{D}}\odot \widehat{\mathbf{H}}\right).
\end{equation}
Next, we analyze the asymptotic behavior of $\mathbf{M}_\mathrm{C}$ and $\mathbf{M}_\mathrm{D}$.

\subsection{Asymptotic Analysis}
\label{subsec:analysis_large_MIMO}
We first extend the law of large numbers for very long random vectors with \ac{i.n.i.d.} entries, to the more general case of very long random vectors with {\it weighted} \ac{i.n.i.d.} entries, where a condition is imposed onto the weights to guarantee the convergence, as summarized in the following lemma.
\begin{lemma}
\label{lemma:long-vector}
Let $\mathbf{p} \triangleq \left[a_1^{\frac{1}{2}} \, p_1, a_2^{\frac{1}{2}} \, p_2, \ldots, a_n^{\frac{1}{2}} \, p_n\right]^{\rm T}$ and $\mathbf{q} \triangleq \left[b_1^{\frac{1}{2}} \, q_1, b_2^{\frac{1}{2}} \, q_2, \ldots, b_n^{\frac{1}{2}} \, q_n\right]^{\rm T}$ be $n \times 1$ vectors, where $a_i, \, b_i \in \mathcal{R}$ are constant coefficients whereas $p_i$ and $q_i$ are \ac{i.n.i.d.} zero-mean \acp{RV}, with $\mathbb{E}\left\{|p_i|^2\right\}=\mu_{p, i}$, $\mathbb{E}\left\{|q_i|^2\right\}=\mu_{q, i}$, $\mathrm{Var}\left\{|p_i|^2\right\}=\sigma^2_{p, i} < \infty$, $\mathrm{Var}\left\{|q_i|^2\right\}=\sigma^2_{q, i} < \infty$, for all $i \in \left\{1,2,\cdots, n\right\}$. If there exists a constant $z < \infty$ such that
\begin{equation}
\label{app_eq:long-vector-1_condition}
|a_i|^2 \leq z,~\forall\, i \in \left\{1,2,\cdots, n\right\},
\end{equation}
then, we have
\begin{equation}
\label{app_eq:long-vector_convergence-1}
\lim_{n \to \infty}\frac{1}{n}\mathbf{p}^{\rm H}\mathbf{p} \overset{p.} \longrightarrow \frac{1}{n} \sum\limits_{k=1}^{n}a_{k} \, \mu_{p, k}
\end{equation}
and
\begin{equation}
\label{app_eq:long-vector_convergence-2}
\lim_{n \to \infty}\frac{1}{n}\mathbf{p}^{\rm H}\mathbf{q} \overset{p.} \longrightarrow 0,
\end{equation}
where $\overset{p.}\longrightarrow$ denotes the convergence in probability.
\end{lemma}
\begin{IEEEproof}
See Appendix~\ref{app_sec:Proof_of_lemma_long-vector}.
\end{IEEEproof}

In particular, Lemma \ref{lemma:long-vector} generalizes the results shown in Eqs. (4) and (5) of \cite{Hien2013} and those in (6) and (7) of \cite{Yang2014}. The results of \cite{Hien2013} apply only to very long random vectors with \ac{i.i.d.} elements while those of \cite{Yang2014} adapt to very long random vectors with \ac{i.n.i.d.} entries. The above Lemma \ref{lemma:long-vector}, on the other hand, is suitable for very long random vectors with {\it weighted} \ac{i.n.i.d.} elements, where the weights must be subject to \eqref{app_eq:long-vector-1_condition}. It is noted that Eqs. (6) and (7) of \cite{Yang2014} were removed from its final version \cite{Yang2015}.

By virtue of Lemma~\ref{lemma:long-vector} and after performing some lengthy algebraic manipulations, the asymptotic behavior of $\mathbf{M}_\mathrm{C}$ and $\mathbf{M}_\mathrm{D}$, defined respectively in \eqref{eq:M_C} and \eqref{eq:M_D}, are attained and summarized in the following theorem.

\begin{theorem}
\label{theorem:asymp_behav_diagonal_of_Mc_and_Md}
If $N_R \to \infty$, $\mathbf{M}_\mathrm{C}$ and $\mathbf{M}_\mathrm{D}$ defined respectively in \eqref{eq:M_C} and \eqref{eq:M_D} become diagonal matrices, and their asymptotic behavior is uniformly given by
\begin{eqnarray}
\label{eq:asymp_behav_diagonal_of_M}
\mathbf{M}_\mathrm{J}
&\overset{p.}\longrightarrow \mathrm{diag}&\left\{\sum\limits_{k=1}^{N_R}d_k^{-\nu}\lambda_{T, 1}\Omega,~\sum\limits_{k=1}^{N_R}d_k^{-\nu} \lambda_{T, 2}\Omega,\right. \nonumber \\
&          & {}\left.~\ldots,~\sum\limits_{k=1}^{N_R}d_k^{-\nu}\lambda_{T, N_T}\Omega   \right\},
\end{eqnarray}
where $ \mathrm{J} \in \{\mathrm{C}, \mathrm{D}\}$, and $d_k \equiv d$,~$\forall k \in \{1, 2,\ldots,N_R\}$, in case of $\mathrm{J}=\mathrm{C}$.
\end{theorem}
\begin{IEEEproof}[Proof]
See Appendix~\ref{app:proof_theorem_asymp_behav_diagonal_of_Mc_and_Md}.
\end{IEEEproof}

Now, by virtue of Theorem \ref{theorem:asymp_behav_diagonal_of_Mc_and_Md}, we can analyze the asymptotic behavior of the spectral efficiencies of both \ac{C-MIMO} and \ac{D-MIMO}, as summarized in the following theorem.
\begin{theorem}
\label{theorem:ergodic-capa_C-MIMO_and_D-MIMO}
Denote the spectral efficiency of \ac{C-MIMO} as $\overline{C}_\mathrm{C}$ and of \ac{D-MIMO} as $\overline{C}_\mathrm{D}$. Their asymptotic behaviors are given by
\begin{equation}
\label{eq:ergodic_capacity_C-MIMO}
\overline{C}_\mathrm{C} \overset{p.}\longrightarrow \sum_{i=1}^{N_T} \log_{2}\left(1+A_iN_Rd^{-\nu}\right)
\end{equation}
and
\begin{equation}
\label{eq:ergodic_capacity_D-MIMO}
\overline{C}_\mathrm{D} \overset{p.}\longrightarrow \sum_{i=1}^{N_T} \log_{2}\left(1+A_i\sum\limits_{k=1}^{N_R}d_k^{-\nu}\right),
\end{equation}
respectively, where $A_i \triangleq \ \frac{\rho}{N_T}\lambda_{T,i}\Omega$,~$\forall i \in \{1, 2,\ldots,N_T\}$.
\end{theorem}
\begin{IEEEproof}[Proof]
By recalling that the asymptotic forms of $\mathbf{M}_\mathrm{C}$ and $\mathbf{M}_\mathrm{D}$ are diagonal matrices explicitly shown in \eqref{eq:asymp_behav_diagonal_of_M}, and by inserting \eqref{eq:asymp_behav_diagonal_of_M} into \eqref{eq:capacity_definition_C-MIMO-1} for \ac{C-MIMO} (with $d_k \equiv d$, for all $k \in \{1, 2,\ldots,N_R\}$) and into \eqref{eq:capacity_definition_D-MIMO-1} for \ac{D-MIMO}, the intended \eqref{eq:ergodic_capacity_C-MIMO} and \eqref{eq:ergodic_capacity_D-MIMO} can be readily obtained, respectively.
\end{IEEEproof}

Based on \eqref{eq:ergodic_capacity_C-MIMO} and \eqref{eq:ergodic_capacity_D-MIMO}, several illuminating insights into the system performance can be immediately gained as follows.
\begin{enumerate}
\item To begin with, \eqref{eq:ergodic_capacity_C-MIMO} and \eqref{eq:ergodic_capacity_D-MIMO} disclose that, in terms of the spectral efficiency of massive \ac{MIMO}, the \ac{D-MIMO} scheme does not always outperform \ac{C-MIMO}. To demonstrate this, let $\delta\triangleq \sum_{k=1}^{N_R}d_k^{-\nu}$. Then, $\overline{C}_\mathrm{D}$ in \eqref{eq:ergodic_capacity_D-MIMO} can be concisely rewritten as $\overline{C}_\mathrm{D} \overset{p.}\longrightarrow \sum_{i=1}^{N_T} \log_{2}\left(1+A_i\delta\right)$. Thus, it is evident that
\begin{itemize}
\item If $N_Rd^{-\nu} < \delta$, we have $\overline{C}_\mathrm{C} < \overline{C}_\mathrm{D}$;
\item If $N_Rd^{-\nu} = \delta$, then $\overline{C}_\mathrm{C} = \overline{C}_\mathrm{D}$, and
\item If $N_Rd^{-\nu} > \delta$, $\overline{C}_\mathrm{C} > \overline{C}_\mathrm{D}$.
\end{itemize}
Although \ac{D-MIMO} does not always outperform \ac{C-MIMO} in terms of spectral efficiency as illustrated above, \ac{D-MIMO} exhibits higher diversity and multiplexing gains. This can be observed from \eqref{eq:ergodic_capacity_D-MIMO} which shows that in the distributed setting, the different path losses over different links between the \ac{BS} antennas and the \ac{UE}, offer additional macro-diversity to the \ac{D-MIMO}. Subsequently, the \ac{D-MIMO} multiplexing gain is up to $N_R N_T$ as shown by \eqref{eq:ergodic_capacity_D-MIMO} whereas that of \ac{C-MIMO} is only $\min(N_T, N_R)$. By recalling that $N_R \to \infty$ in massive \ac{MIMO} context, this result shows that the multiplexing gain is very large in the distributed setting. Consequently, \ac{D-MIMO} is a better choice over \ac{C-MIMO}, to increase the diversity and the multiplexing gains of massive \ac{MIMO} in practice. In particular, with a single-antenna \ac{UE}, i.e., when $N_T=1$, there is neither diversity nor multiplexing gain in case \ac{C-MIMO} is concerned as shown by \eqref{eq:ergodic_capacity_C-MIMO}, whereas in the \ac{D-MIMO} setting the multiplexing gain is $N_R$. Therefore, path-loss factors are crucial to realize diversity and multiplexing gains in distributed schemes.

\item As already known in the classical \ac{C-MIMO} systems, due to insufficient antenna spacing, antenna correlation can significantly degrade the performance of massive \ac{MIMO} systems and has been a subject of practical measurement campaigns (see e.g., \cite{Hoydis2013} and references therein). More specifically, by assuming that there is no correlation at the \ac{BS} side under the \ac{D-MIMO} setting, Eqs. \eqref{eq:ergodic_capacity_C-MIMO} and \eqref{eq:ergodic_capacity_D-MIMO} reveal that the asymptotic behavior of \ac{C-MIMO} and \ac{D-MIMO} are related to the eigenvalues at the \ac{UE} side, i.e., $\lambda_{T,i}$, which shows that the uplink performance of massive \ac{MIMO} in either the centralized setting or the distributed one is determined by the correlation at the \ac{UE} side, given that the Weichselberger correlation model is applied. In particular, as implied by Eqs. \eqref{eq:ergodic_capacity_C-MIMO} and \eqref{eq:ergodic_capacity_D-MIMO}, good channels (i.e. the channels with lower correlation and then larger asymptotic spectral efficiency) are characterized by higher values of $\lambda_{T, i}$, for all $i \in [1, N_T]$, whereas bad channels are characterized by lower values of $\lambda_{T, i}$, for all $i \in [1, N_T]$.

\item By setting $d_{k}=d$ for all $k \in  \{1, 2,\ldots,N_R\}$ in \eqref{eq:ergodic_capacity_D-MIMO}, the macro-diversity gain of \ac{D-MIMO} is degraded and \eqref{eq:ergodic_capacity_D-MIMO} reduces to \eqref{eq:ergodic_capacity_C-MIMO}. Also, from \eqref{eq:ergodic_capacity_C-MIMO} and \eqref{eq:ergodic_capacity_D-MIMO}, it is clear that the spectral efficiency increases with better shadowing conditions (i.e., larger values of $\Omega$ in \eqref{eq:ergodic_capacity_C-MIMO} and \eqref{eq:ergodic_capacity_D-MIMO}), as expected. The same conclusion can be drawn regarding the path loss where, a decrease in the path-loss exponent $\nu$ in \eqref{eq:ergodic_capacity_C-MIMO} and \eqref{eq:ergodic_capacity_D-MIMO} (respectively a decreasing in the distance $d$ in \eqref{eq:ergodic_capacity_C-MIMO} and/or the distances $d_k$ in \eqref{eq:ergodic_capacity_D-MIMO}), benefits improving the spectral efficiency.
\end{enumerate}

In the next subsection, by considering the range of the medium-to-high SNR segment (i.e., $\rho \gg 0$ and even $\rho \to \infty$), \eqref{eq:ergodic_capacity_C-MIMO} and \eqref{eq:ergodic_capacity_D-MIMO} can be further simplified so as to gain more penetrating insights into the spectral efficiency for both the centralized and the distributed schemes.

\subsection{Analysis with Respect to Medium-to-High SNR}
\label{subsec:medium_and_high_SNR_analysis}

In the medium-to-high SNR regime, by recalling that $N_R \to \infty $, \eqref{eq:ergodic_capacity_C-MIMO} and \eqref{eq:ergodic_capacity_D-MIMO} can be further simplified, as summarized in the following corollary.
\begin{corollary}
\label{corrolary:ergodic-capa_C-MIMO_and_D-MIMO-high-SNR-1}
In the medium-to-high \ac{SNR} regime, the asymptotic behavior of the spectral efficiency of the \ac{C-MIMO} shown in \eqref{eq:ergodic_capacity_C-MIMO}, and of the \ac{D-MIMO} shown in \eqref{eq:ergodic_capacity_D-MIMO}, are given by
\begin{equation}
\label{eq:ergodic_capacity_C-MIMO-high-SNR-1}
\overline{C}_{C,\,\mathrm{SNR}\uparrow} \overset{p.}\longrightarrow \sum\limits_{i=1}^{N_T} \log_{2}\lambda_{T,i}+N_T \log_{2}\left({\frac{\rho \, \Omega \, N_R}{N_T}{ d^{-\nu}}}\right)
\end{equation}
and
\begin{equation}
\label{eq:ergodic_capacity_D-MIMO-high-SNR-1}
\overline{C}_{D,\,\mathrm{SNR}\uparrow} \overset{p.}\longrightarrow \sum\limits_{i=1}^{N_T} \log_{2}\lambda_{T,i}+N_T \log_{2}\left(\frac{\rho \, \Omega}{N_T}\sum\limits_{k=1}^{N_R}d_k^{-\nu}\right).
\end{equation}
respectively.
\end{corollary}
\begin{IEEEproof}[Proof]
By applying the identity $\log_2(1+x) \approx \log_2(x)$ if $x\to \infty$ to \eqref{eq:ergodic_capacity_C-MIMO} and \eqref{eq:ergodic_capacity_D-MIMO}, Eqs. \eqref{eq:ergodic_capacity_C-MIMO-high-SNR-1} and \eqref{eq:ergodic_capacity_D-MIMO-high-SNR-1} can be readily obtained.
\end{IEEEproof}

From \eqref{eq:ergodic_capacity_C-MIMO-high-SNR-1} and \eqref{eq:ergodic_capacity_D-MIMO-high-SNR-1}, it is evident that:
\begin{enumerate}
\item The asymptotic spectral efficiency increases {\it logarithmically} with the average SNR ($\rho$), the average power of the shadowing effect ($\Omega$), and the number of Rx antennas ($N_R$), in the medium-to-high SNR regime;

\item The asymptotic spectral efficiency decreases exponentially with the distance ($d$) and the path-loss exponent ($\nu$), and increases with both the number of Tx ($N_T$) and the number of Rx ($N_R$) antennas.
\end{enumerate}

The results obtained in the above can be applied to \ac{C-MIMO} and \ac{D-MIMO} where the cells are designed with arbitrary topology, for instance, hexagonal topology, circular topology, line topology and grid topology. In the next section, a case study is performed by applying the obtained results into circular topology, which is widely used in the open literature for performance evaluation of various wireless systems \cite{Xinzheng2009, Firouzabadi2011, Dongming2013}. In particular, circular topology is an excellent setting for performance analysis of massive \ac{MIMO} where, by taking \ac{D-MIMO} for instance, the optimization of antennas' locations becomes very challenging due to the large number of antennas and the complex system parameters involved.

\section{Application to Circular Network}
\label{sec:location_optimization}

In this part, a case study where the network adopts circular topology is investigated. After detailing the system model, the asymptotic spectral efficiency for both \ac{C-MIMO} and \ac{D-MIMO} schemes is derived. Then, the average spectral efficiency for a user is attained, by assuming a uniform user distribution and a typical urban propagation environment. Finally, the optimal antenna deployment pertaining to D-MIMO is studied.

\subsection{System Model for the Circular Topology}
\label{subsec:system_circular}

\begin{figure}[t]
\centering
\includegraphics[width=3.5in]{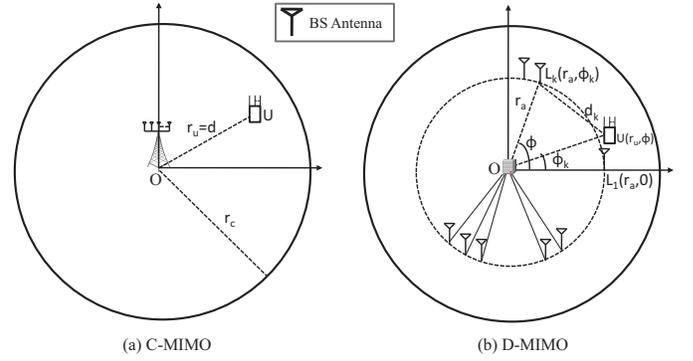}
\caption{C-MIMO and D-MIMO systems with circular topology. Coverage areas are of circular shape with radius $r_\mathrm{c}$, and the antennas of D-MIMO are uniformly distributed along a circle with radius $r_\mathrm{a}$.}
\label{fig:system_model_circular}
\end{figure}

As depicted in Fig.~\ref{fig:system_model_circular}, we consider a point-to-point uplink \ac{MIMO} system, where the \ac{BS} is assumed to be centered at the origin $(0, 0)$ and its circular coverage area is of radius $r_{\mathrm{c}}$. In the \ac{C-MIMO} setting shown in Fig.~\ref{fig:system_model_circular}-a, the receive antennas are all co-located at the \ac{BS}, whereas in the \ac{D-MIMO} system shown in Fig.~\ref{fig:system_model_circular}-b, the \ac{BS} antennas are uniformly deployed along a circle of radius $r_{\mathrm{a}}$, which is concentric with the circular cell of radius $r_{\mathrm{c}}$. In the latter configuration, for the $k^{\mathrm{th}}$ \ac{BS} antenna, $k \in \{1, 2,\ldots,N_R\}$, its location is denoted as $L_k$, specifying its polar coordinates relative to the center of the coverage area denoted by $(r_{\mathrm{a}},\phi_k)$, where $\phi_k \in [0, 2\pi)$. In both \ac{MIMO} schemes, the location of the user is denoted as $U$. Moreover, in the centralized setting the distance between the \ac{UE} and the \ac{BS} is given by $d$, while in the distributed scheme the polar coordinates of the user are denoted as $(r_{\mathrm{u}},\phi)$, with $r_{\mathrm{u}}$ being the distance between the \ac{UE} and the cell center, and with $\phi \in [0, 2\pi)$. Without loss of generality, the $N_R$ \ac{BS} Rx antennas are assumed to form a uniform circular antenna array with $\phi_1 = 0$ and $\phi_k = 2\pi(k-1)/{N_R}$, for all $ k \in [2, N_R]$, as shown in Fig.~\ref{fig:system_model_circular}-b. On the other hand, to keep consistent with the parameters used in the previous sections, the distance between the \ac{UE} and the $k^{\mathrm{th}}$ antenna of the \ac{BS} pertaining to  \ac{D-MIMO} is given by $d_k$, for all $k \in \{1, 2,\ldots,N_R\}$. With these settings in mind, in the next subsection we derive the asymptotic spectral efficiency.

\subsection{Asymptotic Spectral Efficiency in Circular Topology}
\label{subsec:asymptotic_behavior_circular}

Since the \ac{BS} antennas are co-located in the \ac{C-MIMO} scheme, the spectral efficiency is similar to \eqref{eq:ergodic_capacity_C-MIMO}, where $d$ is replaced by $r_{\mathrm{u}}$. In the \ac{D-MIMO} setting, however, the asymptotic spectral efficiency is derived by exploiting the law of cosines, and the ensuing results are summarized in the following theorem.
\begin{theorem}
\label{theorem:ergodic-capa_C-MIMO_and_D-MIMO_circular}
For a UE located at distance $r_{\mathrm{u}}$ from the cell center of the massive \ac{C-MIMO} and \ac{D-MIMO} systems of circular topology, as illustrated in Fig.~\ref{fig:system_model_circular}, the asymptotic spectral efficiencies are given by
\begin{equation}
\label{eq:ergodic_capacity_C-MIMO_circular}
\overline{C}_\mathrm{C, \, circ} \xrightarrow{p.} \sum\limits_{i=1}^{N_T} \log_{2}\left(1+\beta_{i} \, r_{\mathrm{u}}^{-\nu}\right)
\end{equation}
and
\begin{equation}
\label{eq:ergodic_capacity_D-MIMO_circular}
\overline{C}_\mathrm{D, \, circ}\xrightarrow{p.}\sum\limits_{i=1}^{N_T} \log_{2}\left[1+\beta_{i} \, \left|r_{\mathrm{u}}^{2}-r_{\mathrm{a}}^{2}\right|^{\frac{\nu}{2}}P_{\frac{\nu}{2}-1}\left(\frac{r_{\mathrm{u}}^{2}+r_{\mathrm{a}}^{2}}{\left|r_{\mathrm{u}}^{2}-r_{\mathrm{a}}^{2}\right|}\right)\right],
\end{equation}
respectively, where $\beta_{i} \triangleq N_R A_i$ and $P_{c}(x)$ refers to the Legendre function of the first kind \cite[Eq. (8.820)]{Gradshteyn2007}.
\end{theorem}
\begin{IEEEproof}[Proof]
Equation~\eqref{eq:ergodic_capacity_C-MIMO_circular} is immediately available by replacing $d$ in \eqref{eq:ergodic_capacity_C-MIMO} with $r_u$. On the other hand, the proof of \eqref{eq:ergodic_capacity_D-MIMO_circular} is provided in Appendix \ref{Proof_of_equation_ergodic_capacity_D-MIMO_circular}.
\end{IEEEproof}

From \eqref{eq:ergodic_capacity_C-MIMO_circular} and \eqref{eq:ergodic_capacity_D-MIMO_circular}, similar insights to those presented after Theorem~\ref{theorem:ergodic-capa_C-MIMO_and_D-MIMO} can be gained. Also, by following similar procedure as in Section \ref{subsec:medium_and_high_SNR_analysis}, similar results to those in Section \ref{subsec:medium_and_high_SNR_analysis} can be obtained. More importantly, \eqref{eq:ergodic_capacity_C-MIMO_circular} and \eqref{eq:ergodic_capacity_D-MIMO_circular} can be numerically computed and serve as performance benchmark in various propagation environments. In the next subsection, we consider a typical urban environment and random values of $r_{\mathrm{u}}$, and derive asymptotic expression of the average spectral efficiency (w.r.t. $N_R$) for a typical user in the cell, which reflects the achievable data rate over a normalized spectral bandwidth.

\subsection{Average Asymptotic Spectral Efficiency in Urban Area}
\label{subsec:average_asymptotic_behavior_circular}
Assuming transmission in an urban area with path-loss exponent $\nu=4$, for a randomly deployed UE, i.e., the value of $r_{\mathrm{u}}$ is random, we derive the average asymptotic spectral efficiency of such a typical user in the cell, in both centralized and distributed settings. We assume that $r_{\mathrm{u}}$ is uniformly distributed in $(0, r_{\mathrm{c}}]$. Then, the \ac{PDF} of $r_{\mathrm{u}}$ is given by
\begin{equation}
\label{eq:r_u_distribution}
 f_{r_{\mathrm{u}}}(x) = \left\{
  \begin{array}{l l}
    \frac{2}{r_{\mathrm{c}}^{2}}x, & \quad \text{if $0 < x \leq r_{\mathrm{c}}$;} \\
    0, & \quad \text{otherwise}.
  \end{array} \right.
\end{equation}

In case of $\nu=4$, the asymptotic behaviors of \ac{C-MIMO} shown in \eqref{eq:ergodic_capacity_C-MIMO_circular} and that of \ac{D-MIMO} given by \eqref{eq:ergodic_capacity_D-MIMO_circular}, reduce to
\begin{equation}
\label{eq:particular_values_of_capacity_C-MIMO_circular_Legendre_P-4}
\overline{C}_\mathrm{C, \, circ} \xrightarrow{p.} \sum\limits_{i=1}^{N_T} \log_{2}\left(1+\frac{\beta_{i}}{r_{\mathrm{u}}^{4}}\right)
\end{equation}
and
\begin{equation}
\overline{C}_\mathrm{D, \, circ}\xrightarrow{p.}
\label{eq:particular_values_of_capacity_D-MIMO_circular_Legendre_P-4}
\sum\limits_{i=1}^{N_T} \log_{2}\left[1+\beta_i\frac{r_{\mathrm{u}}^{2}+r_{\mathrm{a}}^{2}}{|r_{\mathrm{u}}^{2}-r_{\mathrm{a}}^{2}|^{3}}\right],
\end{equation}
respectively.

From \eqref{eq:particular_values_of_capacity_C-MIMO_circular_Legendre_P-4} and \eqref{eq:particular_values_of_capacity_D-MIMO_circular_Legendre_P-4}, it is evident that, by setting $r_{\mathrm{a}}\to 0$, \eqref{eq:particular_values_of_capacity_D-MIMO_circular_Legendre_P-4} reduces to \eqref{eq:particular_values_of_capacity_C-MIMO_circular_Legendre_P-4}. Therefore, in the following we derive only the average asymptotic spectral efficiency in the \ac{D-MIMO} setting w.r.t. $r_u$. That of the \ac{C-MIMO} scheme can be readily attained by setting $r_{\mathrm{a}}\to 0$. The results are summarized in the following theorem.

\begin{theorem}
\label{theorem:average_asymptotic_behavior_circular_C-MIMO_and_D-MIMO}
The average asymptotic spectral efficiencies of the \ac{C-MIMO} and \ac{D-MIMO} systems in an urban area w.r.t. a uniformly distributed user in the coverage area,  are given by
\begin{equation}
\label{eq:average_asymptotic_behavior_circular_C-MIMO}
 \overline{C}_\mathrm{C, \, circ}^{\mathrm{avg}}
\xrightarrow{p.} 2N_T\log_{2}\frac{e}{r_{\mathrm{c}}^{2}} + \sum\limits_{i=1}^{N_T}\log_{2}{\beta_i}
\end{equation}
and
\begin{eqnarray}
\label{eq:average_asymptotic_behavior_circular_D-MIMO}
\overline{C}_\mathrm{D, \, circ}^{\mathrm{avg}}
& \xrightarrow{p.}    & 2N_T\log_{2}e-4N_T\frac{r_{\mathrm{a}}^{2}}{r_{\mathrm{c}}^{2}}\log_{2}r_{\mathrm{a}}^{2} \nonumber \\
&                             &   -3N_T\left(1-\frac{r_{\mathrm{a}}^{2}}{r_{\mathrm{c}}^{2}}\right)\log_{2}\left(r_{\mathrm{c}}^{2}-r_{\mathrm{a}}^{2}\right) \nonumber \\
&                             &   + N_T\left(1+\frac{r_{\mathrm{a}}^{2}}{r_{\mathrm{c}}^{2}}\right)\log_{2}\left(r_{\mathrm{c}}^{2}+r_{\mathrm{a}}^{2}\right) \nonumber \\
&                             &   {}+ \sum\limits_{i=1}^{N_T}\log_{2}{\beta_i},
\end{eqnarray}
respectively.
\end{theorem}
\begin{IEEEproof}[Proof]
The proof of \eqref{eq:average_asymptotic_behavior_circular_D-MIMO} is provided in Appendix~\ref{Proof_of_average_asymptotic_behavior_circular_C-MIMO_and_D-MIMO}. Equation \eqref{eq:average_asymptotic_behavior_circular_C-MIMO} can be easily derived by setting $r_{\mathrm{a}}\to 0$ in \eqref{eq:average_asymptotic_behavior_circular_D-MIMO}.
\end{IEEEproof}

From \eqref{eq:average_asymptotic_behavior_circular_D-MIMO}, it is clear that the average spectral efficiency varies with the radius of antenna array, $r_{\mathrm{a}}$, given that the coverage radius $r_{\mathrm{c}}$ is fixed. Therefore, in the next subsection, we derive the optimal value of $r_{\mathrm{a}}$ that maximizes the average spectral efficiency given by \eqref{eq:average_asymptotic_behavior_circular_D-MIMO}.

\subsection{Optimal Location of the Antenna Array in the \ac{D-MIMO} Setting}
\label{subsec:optimal_location_antenna_array_D-MIMO}

After some lengthy yet straightforward algebraic manipulations, the optimal value of $r_a$ that maximizes the average spectral efficiency is discovered and formalized as follows.

\begin{corollary}
\label{corollary:optimal_location_antenna_array_D-MIMO}
For a massive \ac{D-MIMO} system operating in an urban area, with cell radius $r_{\mathrm{c}}$, the optimal value of the radius of the circular antenna array that maximizes the average spectral efficiency given by \eqref{eq:average_asymptotic_behavior_circular_D-MIMO} is determined by
\begin{equation}
\label{eq:optimal_antenna_array_dimension}
r_{\mathrm{a}}^{\mathrm{opt}}=r_{\mathrm{c}}/1.31.
\end{equation}
\end{corollary}
\begin{IEEEproof}[Proof]
See Appendix \ref{Proof_of_corollary:optimal_location_antenna_array_D-MIMO}.
\end{IEEEproof}

It is very interesting to see from \eqref{eq:optimal_antenna_array_dimension} that the optimal value of the antenna array size which maximizes the average spectral efficiency given by \eqref{eq:average_asymptotic_behavior_circular_D-MIMO}, is independent of $\beta_i$ or $\lambda_{T,i}$, for all $i \in \{1, 2,\ldots,N_T\}$. Thus, the optimal antenna radius in the \ac{D-MIMO} setting depends neither upon the shadowing parameters nor the correlation factors, even if the shadowing and antenna correlation have severe effects on the performance of massive \ac{MIMO} systems. Moreover, \eqref{eq:optimal_antenna_array_dimension} implies that the optimal value of the antenna array size is independent of the average SNR and of the numbers of Tx/Rx antennas. These findings shed new light on the design and deployment of massive \ac{D-MIMO} systems in practice.
\section{Numerical Results and Discussions}
\label{sec:simulations}

\begin{figure}[t]
\centering
\includegraphics[width=3.5in,height=2.7in]{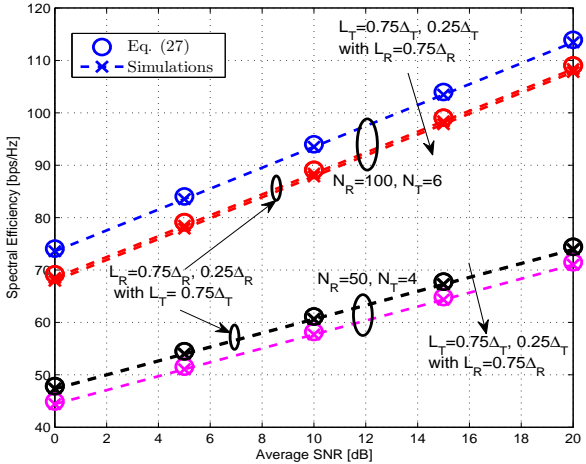}
\caption{The asymptotic spectral efficiency of \ac{C-MIMO} w.r.t. the antenna spacing ($d=0.2$ and $\nu=3.7$).}
\label{fig:Correlation_C_MIMO}
\end{figure}

\begin{figure}[t]
\centering
\includegraphics[width=3.5in,height=2.7in]{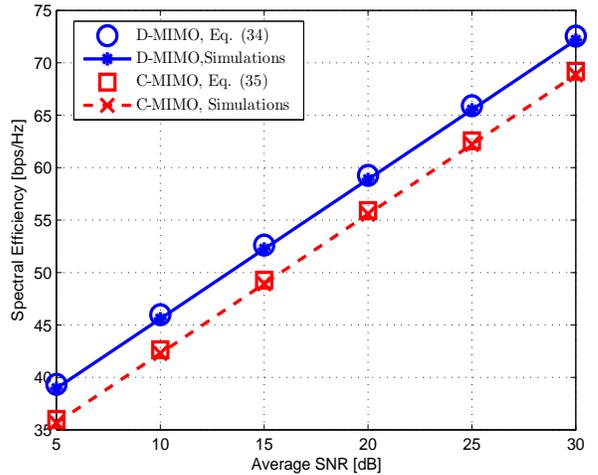}
\caption{The asymptotic spectral efficiency of \ac{C-MIMO} and \ac{D-MIMO} with circular topology ($r_{\mathrm{a}}=0.2$, $r_{\mathrm{u}}=0.5$, $\nu=3.7$, $N_T=4$ and $N_R=100$).}
\label{fig:C-MIMO_D_MIMO_Circular}
\end{figure}

In this section, we present and discuss ensuing simulations results, compared with numerical ones pertaining to the analysis developed previously. The simulation experiments, of Monte Carlo type, are performed in the platform Matlab R2014a. In the simulation setting, the variance of AWGN at \ac{UE} ($N_{0}$) is set to unity and, unless otherwise stated, the mean local power of the shadowing effect ($\Omega$) is normalized w.r.t. $N_{0}$ and set to $\Omega=0$ dB.  Apart from Fig.~\ref{fig:Correlation_C_MIMO}, the sub-array spacing is set to $L_T=0.25 {\Delta_T}$ and $L_R= 0.75 {\Delta_R}$. The spectral efficiency is in the unit of bit/s/Hz and the average \ac{SNR} is in the unit of dB w.r.t. $N_{0}$. Regarding the number of Rx antennas at the \ac{BS}, $N_R$, although no standard value has yet been specified for practical massive \ac{MIMO} deployments, in densely populated areas such as stadiums where a \ac{BS} may serve thousands of UEs, one can imagine that $N_R$ may be equal to, or even greater than, $100$ or $200$, whereas for areas with few UEs, $N_R$ may take smaller values such as $20$. Below, in different simulation scenarios, the value of $N_R$ ranges between $20$ and $200$. In Figs.~\ref{fig:C-MIMO_D_MIMO_Circular}-\ref{fig:D-MIMO_Circular_Average_opt_Ra} where circular topology is adopted, the radius of circular coverage area is set to $r_{\mathrm{c}}=500$ m, which is the typical value for cellular cells in up-to-date cellular networks. Finally, in all the following simulation scenarios, distance parameters are normalized w.r.t. a reference $R_0=500$~m. Such normalization is widely adopted in the related literature, see e.g., \cite{Hoydis2013}.

In the following, we first discuss the results pertaining to the asymptotic spectral efficiency for both centralized and distributed schemes, developed in Section~\ref{sec:asymptotic_capacity}.

\begin{figure}[t]
\centering
\includegraphics[width=3.5in,height=2.7in]{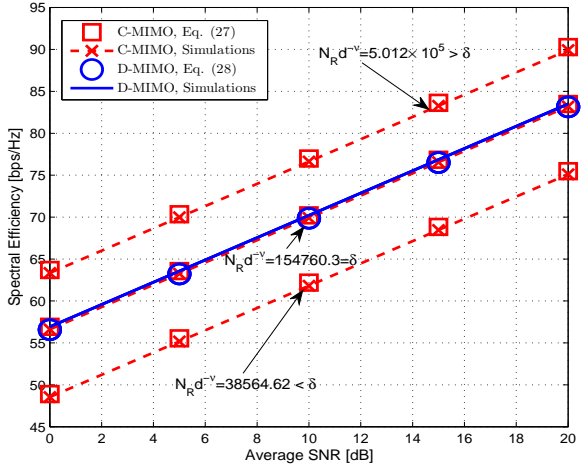}
\caption{The asymptotic spectral efficiency of \ac{C-MIMO} and \ac{D-MIMO} ($\nu=3.7$, $\delta=154760.3$, $N_R=100$ and $N_T=4$).}
\label{fig:ergodic_capacity_Comparison_C_MIMO_D_MIMO}
\end{figure}

\subsection{Spectral-Efficiency Comparison Between \ac{C-MIMO} and \ac{D-MIMO} Schemes}
\label{sim_subsec:ergodic_capacity_comparison_C_MIMO_D_MIMO}

Figure~\ref{fig:ergodic_capacity_Comparison_C_MIMO_D_MIMO} compares the spectral efficiency of massive \ac{C-MIMO} and \ac{D-MIMO} systems. By recalling that $\delta \triangleq \sum_{k=1}^{N_R}d_k^{-\nu}$ as defined after \eqref{eq:ergodic_capacity_D-MIMO} and by using the normalization w.r.t. $R_0$, we set, for simulation purpose, $d_k=R_0^{-1}(k+100)$, for all $k \in \{1, 2,\ldots,N_R\}$, $\nu=3.7$, $N_R=100$, $N_T=4$, and vary the value of $d$ shown in \eqref{eq:ergodic_capacity_C-MIMO}. With this setting, it is easy to get $\delta=154760.3$. It is observed from the curves in the middle of Fig.~\ref{fig:ergodic_capacity_Comparison_C_MIMO_D_MIMO} that $\overline{C}_\mathrm{C}=\overline{C}_\mathrm{D}$ in case of $\delta = N_Rd^{-\nu}$. The lower curves of Fig.~\ref{fig:ergodic_capacity_Comparison_C_MIMO_D_MIMO} show that $\overline{C}_\mathrm{C}<\overline{C}_\mathrm{D}$ in case of $N_Rd^{-\nu} < \delta$, whereas the upper curves illustrate that $\overline{C}_\mathrm{C} > \overline{C}_\mathrm{D}$ in case of $N_Rd^{-\nu}  > \delta$. These observations corroborate the results obtained in Section \ref{subsec:analysis_large_MIMO}.

\begin{figure}[t]
\centering
\includegraphics[width=3.5in,height=2.7in]{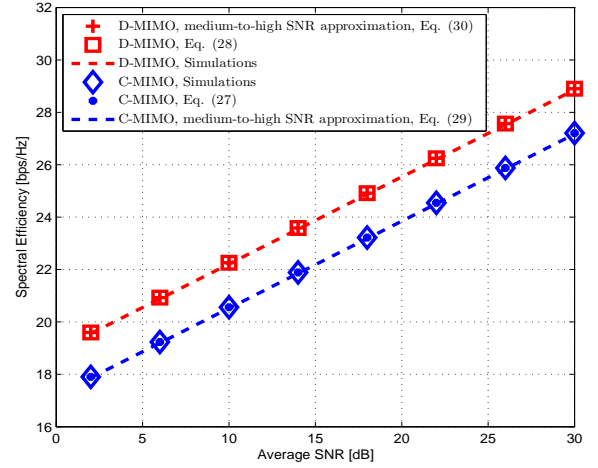}
\caption{The asymptotic spectral efficiency of \ac{C-MIMO} and \ac{D-MIMO} w.r.t. the medium-to-high \ac{SNR} ($d_k=R_0^{-1}(k+100)$, for all $k \in \{1, 2,\ldots,N_R\}$, $d=0.1$, $N_T=1$, $N_R=200$ and $\nu=3.7$).}
\label{fig:High_SNR_C_MIMO_D_MIMO}
\end{figure}

\subsection{Impact of Correlation on the Asymptotic Spectral Efficiency}
\label{sim_subsec:impact_of_correlation}

By recalling that in the \ac{D-MIMO} scheme there is no correlation among antennas at the \ac{BS} side, it is disclosed in Section~\ref{subsec:analysis_large_MIMO} that the system performance of massive \ac{MIMO} in both centralized and distributed settings is determined by the correlation at the \ac{UE} side, in case the Weichselberger correlation model is applied. This is corroborated by observations from Fig.~\ref{fig:Correlation_C_MIMO} where, for sake of clarity, only the curves related to \ac{C-MIMO} are plotted. Clearly, Fig.~\ref{fig:Correlation_C_MIMO} shows that by decreasing the spacing between antennas (then increasing correlation impact) at the \ac{UE} side (the transmitter), the spectral efficiency of massive \ac{MIMO} decreases significantly. However, changing the spacing between antennas at the \ac{BS} side (the receiver), does not essentially affect the system performance. This result corroborates the previous analysis.

\begin{figure}[t]
\centering
\includegraphics[width=3.5in,height=2.7in]{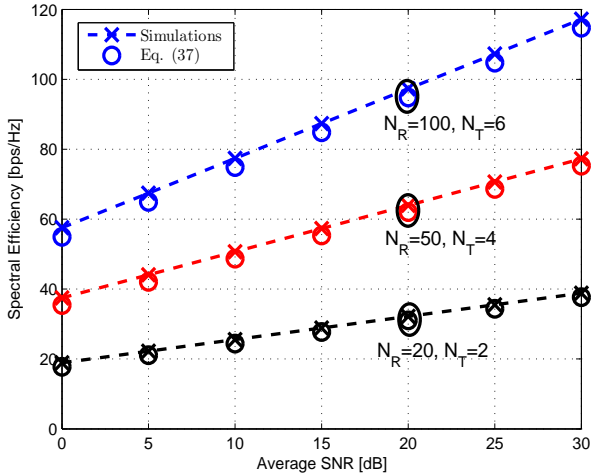}
\caption{The average asymptotic spectral efficiency of \ac{D-MIMO} with circular topology ($\nu=4$, $r_{\mathrm{a}}=0.5$).}
\label{fig:D-MIMO_Circular_Average}
\end{figure}

\subsection{Special Case: Analysis w.r.t. Medium-to-High SNR}

Figure \ref{fig:High_SNR_C_MIMO_D_MIMO} illustrates the asymptotic spectral efficiency of both massive \ac{C-MIMO} and \ac{D-MIMO} schemes in terms of medium-to-high \ac{SNR}, as developed in Section~\ref{subsec:medium_and_high_SNR_analysis}. It is observed that the approximate asymptotic behavior works perfectly for both antenna settings.

\subsection{A Case Study: Asymptotic Spectral Efficiency for Circularly Topology}
\label{sim_subsec:asymptotic_behavior_circular}

Figure~\ref{fig:C-MIMO_D_MIMO_Circular} presents the asymptotic spectral efficiency of massive \ac{C-MIMO} and \ac{D-MIMO} systems of circular topology. The cell size $r_{\mathrm{c}}$ is normalized to unity, i.e., $r_{\mathrm{c}}=1$, $r_{\mathrm{a}}=0.2$ and $r_{\mathrm{u}}=0.5$ are used in the simulation setting. It is observed from Fig.~\ref{fig:C-MIMO_D_MIMO_Circular} that the simulation results agree perfectly with the numerical results computed as per \eqref{eq:ergodic_capacity_C-MIMO_circular} and \eqref{eq:ergodic_capacity_D-MIMO_circular}, respectively.

Figure~\ref{fig:D-MIMO_Circular_Average} shows the average asymptotic spectral efficiency for a user in a cell under the \ac{D-MIMO} (by recalling that the \ac{C-MIMO} scheme here is just a particular case of the \ac{D-MIMO} by setting $r_{\mathrm{a}} \to 0$), in an urban area with $\nu=4$. The results are shown w.r.t. different numbers of Tx/Rx antennas, with $r_{\mathrm{a}}=0.5$. It is seen that the simulation results agree well with the numerical ones computed as per \eqref{eq:average_asymptotic_behavior_circular_D-MIMO}.

\begin{figure}[t]
\centering
\includegraphics[width=3.5in,height=2.7in]{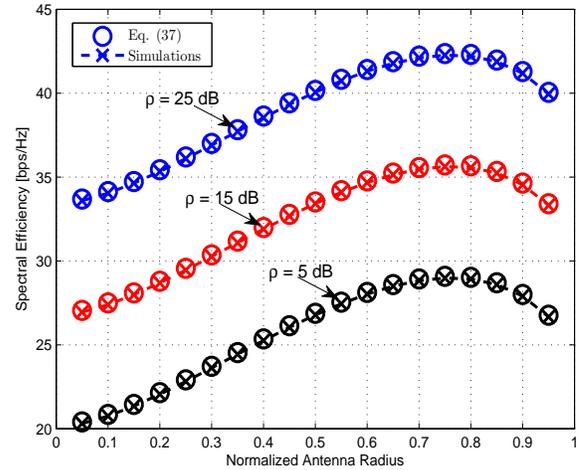}
\caption{Optimal value of the antenna array size that maximizes the average spectral efficiency of \ac{D-MIMO} ($\nu=4$).}
\label{fig:D-MIMO_Circular_Average_opt_Ra}
\end{figure}

Figure \ref{fig:D-MIMO_Circular_Average_opt_Ra} plots the average asymptotic spectral efficiency of massive \ac{D-MIMO} being circular topology w.r.t. the normalized antenna radius, and illustrates the value of the antenna radius which maximizes the average spectral efficiency of the system. It is observed that, for different values of \ac{SNR}, the maximum value of the average spectral efficiency always appears at $r_{\mathrm{a}} \approx 0.76$, which is consistent with our analytical result given by \eqref{eq:optimal_antenna_array_dimension}, i.e., $r_{\mathrm{a}}^{\mathrm{opt}}=r_{\mathrm{c}}/1.31 \approx 0.76$.

\section{Concluding Remarks}
\label{sec:conclusions}

In this paper, the spectral efficiency of massive \ac{MIMO} systems in both centralized (\ac{C-MIMO}) and distributed (\ac{D-MIMO}) settings, was analytically investigated, based on a novel comprehensive analytical channel model where major natural environmental and antenna physical parameters were accounted for, including path loss, shadowing, multi-path fading and antenna correlation. Our results reveal that the \ac{C-MIMO} scheme does not always underperform \ac{D-MIMO}, although the latter exhibits a higher multiplexing gain. Further, the uplink performance of massive \ac{MIMO} is shown to be mainly determined by the antenna correlation at the \ac{UE} side, given that the Weichselberger correlation model is applied. For practical purposes, it was demonstrated that for the \ac{D-MIMO} scheme with circular topology of radius $r_{\mathrm{c}}$, the optimal value of the radius of antenna array that maximizes the average spectral efficiency is given by $r_{\mathrm{a}}^{\mathrm{opt}}=r_{\mathrm{c}}/1.31$. The proposed channel model, the developed analysis and the obtained results, thanks to their generality and compactness, can serve as practical benchmark for designing and analyzing performances of massive MIMO systems in real physical propagation environments.

\appendices

\section{Proof of Lemma \ref{lemma:long-vector}}
\label{app_sec:Proof_of_lemma_long-vector}
With the vectors $\mathbf{p}$ and $\mathbf{q}$ defined in Lemma~\ref{lemma:long-vector}, we have $\mathbf{p}^{\rm H}\mathbf{p}=\sum_{k=1}^{n}a_k|p_k|^2$ and $\mathbb{E}\left\{\mathbf{p}^{\rm H}\mathbf{p}\right\}=\sum_{k=1}^{n}a_{k}\mu_{p, k}$. Let $z$ be a constant which satisfies \eqref{app_eq:long-vector-1_condition}. It is then clear that $\mathrm{Var}\left\{|a_i^{\frac{1}{2}}p_i|^2\right\}=|a_i|^2\sigma^2_{p, i} < \infty$. Therefore, by applying the well-known law of large numbers for \ac{i.n.i.d.} \acp{RV} \cite[p. 313]{Grinstead2012} to $a_i^{\frac{1}{2}}p_i$, the entries of $\mathbf{p}$, \eqref{app_eq:long-vector_convergence-1} is easily attained.

On the other hand, since $p_i$ and $q_i$ are independent, we have $\mathbb{E}\left\{p_i^{\rm H}q_i\right\}=0$, for all $i \in \left\{1,2,\cdots, n\right\}$. Finally, by following similar steps as in the proof of \eqref{app_eq:long-vector_convergence-1}, \eqref{app_eq:long-vector_convergence-2} can be readily proven.

\newcounter{mytempeqncnt1}
\begin{figure*}[!t]
\normalsize
\setcounter{mytempeqncnt1}{\value{equation}}
\setcounter{equation}{39}
\begin{tiny}
\begin{equation}
\label{app-eq:explicit_expression_of_M}
\mathbf{M}_\mathrm{J}=
\begin{bmatrix}
\sum\limits_{k=1}^{N_R} \lambda_{T, 1}\lambda_{R, k}d_k^{-\nu}{\varphi}_k|\widehat{H}_{k, 1}|^{2} & \sum\limits_{k=1}^{N_R} \lambda_{T, 1}^\frac{1}{2}\lambda_{T, 2}^\frac{1}{2}\lambda_{R, k}d_k^{-\nu}{\varphi}_k\widehat{H}_{k, 1}^{\rm H} {\widehat{H}_{k, 2}} & \cdots & \sum\limits_{k=1}^{N_R}\lambda_{T,1}^\frac{1}{2} \lambda_{T, N_T}^\frac{1}{2}\lambda_{R, k}d_k^{-\nu}{\varphi}_k{\widehat{H}_{k,1}}^{\rm H}\widehat{H}_{k,N_T}  \\
\sum\limits_{k=1}^{N_R} \lambda_{T, 2}^\frac{1}{2}\lambda_{T, 1}^\frac{1}{2}\lambda_{R, k}d_k^{-\nu}{\varphi}_k {\widehat{H}_{k, 2}}^{\rm H}\widehat{H}_{k,1} & \sum\limits_{k=1}^{N_R}\lambda_{T, 2} \lambda_{R, k}d_k^{-\nu}{\varphi}_k|\widehat{H}_{k,2}|^{2} & \cdots & \sum\limits_{k=1}^{N_R}\lambda_{T, 2}^\frac{1}{2}\lambda_{T, N_T}^\frac{1}{2} \lambda_{R, k}d_k^{-\nu}{\varphi}_k{\widehat{H}_{k,2}}^{\rm H}\widehat{H}_{k,N_T}  \\
\vdots & \vdots & \ddots & \vdots \\
\sum\limits_{k=1}^{N_R}\lambda_{T, N_T}^\frac{1}{2}\lambda_{T, 1}^\frac{1}{2} \lambda_{R, k}d_k^{-\nu}{\varphi}_k{\widehat{H}_{k,N_T}}^{\rm H}\widehat{H}_{k,1}  & \sum\limits_{k=1}^{N_R}\lambda_{T, N_T}^\frac{1}{2}\lambda_{T, 2}^\frac{1}{2} \lambda_{R, k}d_k^{-\nu}{\varphi}_k{\widehat{H}_{k,N_T}}^{\rm H}\widehat{H}_{k,2}  & \cdots &  \sum\limits_{k=1}^{N_R} \lambda_{T, N_T}\lambda_{R, k}d_k^{-\nu}{\varphi}_k|\widehat{H}_{k,N_T}|^{2}
\end{bmatrix}.
\end{equation}
\end{tiny}
\setcounter{equation}{\value{mytempeqncnt1}}
\hrulefill
\vspace{-10pt}
\end{figure*}
\setcounter{equation}{40}

\newcounter{mytempeqncnt2}
\begin{figure*}[t]
\normalsize
\setcounter{mytempeqncnt2}{\value{equation}}
\setcounter{equation}{44}
\begin{numcases}{\frac{1}{N_R}\mathbf{g}_{i}^{\rm H}\mathbf{h}_{j}\xrightarrow{p.}\frac{1}{N_R}\mathbb{E}\{[\mathbf{M}_\mathrm{J}]_{i, j}\}\overset{p.}\longrightarrow}
\label{app-eq:asymp_behav_of_non_diag_entries_of_Mc}
0, & $i \neq j$;
\\
\label{app-eq:asymp_behav_of_diag_entries_of_Mc}
 \frac{1}{N_R}\sum\limits_{k=1}^{N_R}\lambda_{T, i}\, d_k^{-\nu}\Omega, & $i=j$.
\end{numcases}
\setcounter{equation}{\value{mytempeqncnt2}}
\hrulefill
\vspace{-10pt}
\end{figure*}
\setcounter{equation}{46}

\newcounter{mytempeqncnt3}
\begin{figure*}[!t]
\normalsize
\setcounter{mytempeqncnt3}{\value{equation}}
\setcounter{equation}{46}
\begin{equation}
\label{app-eq:asymp_behav_diagonal_of_Md}
\mathbf{M}_\mathrm{J}=N_R\left(\frac{1}{N_R}\mathbf{M}_\mathrm{D}\right)\overset{p.}\longrightarrow \mathrm{diag}\left\{\sum\limits_{k=1}^{N_R}d_k^{-\nu}\lambda_{T, 1}\Omega,\sum\limits_{k=1}^{N_R}d_k^{-\nu} \lambda_{T, 2}\Omega, \ldots,\sum\limits_{k=1}^{N_R}d_k^{-\nu}\lambda_{T, N_T}\Omega   \right\}.
\end{equation}
\setcounter{equation}{\value{mytempeqncnt3}}
\hrulefill
\vspace{-10pt}
\end{figure*}
\setcounter{equation}{47}

\section{Proof of Theorem \ref{theorem:asymp_behav_diagonal_of_Mc_and_Md}}
\label{app:proof_theorem_asymp_behav_diagonal_of_Mc_and_Md}
We first derive the explicit expressions of matrices $\mathbf{G}_{\mathrm{C}}$ and $\mathbf{G}_{\mathrm{D}}$ uniformly defined in \eqref{eq:eigenmode_coupling_elt_wise_square-root}. By substituting \eqref{eq:singular_value_vectors_Tx} and \eqref{eq:singular_value_vectors_Rx} into \eqref{eq:eigenmode_coupling_elt_wise_square-root} and recalling that $\lambda_{R,k}=1$, for all $k \in \{1, 2,\ldots,N_R\}$, if $\mathrm{J}=\mathrm{D}$ (as $\mathbf{\Theta}_{R}$ reduces to $\mathbf{I}_{N_R}$ in the \ac{D-MIMO} setting), $\mathbf{G}_{\mathrm{C}}$ and $\mathbf{G}_{\mathrm{D}}$ can be uniformly expressed as
\setcounter{equation}{38}
\begin{tiny}
\begin{equation}
\mathbf{G}_{\mathrm{J}}=
  \begin{bmatrix}
\lambda_{R,1}^{\frac{1}{2}}\lambda_{T,1}^{\frac{1}{2}} & \lambda_{R,1}^{\frac{1}{2}}\lambda_{T,2}^{\frac{1}{2}} & \cdots & \lambda_{R,1}^{\frac{1}{2}}\lambda_{T,N_T}^{\frac{1}{2}} \\
\lambda_{R,2}^{\frac{1}{2}}\lambda_{T,1}^{\frac{1}{2}} & \lambda_{R,2}^{\frac{1}{2}}\lambda_{T,2}^{\frac{1}{2}} & \cdots & \lambda_{R,2}^{\frac{1}{2}}\lambda_{T,N_T}^{\frac{1}{2}}  \\
\vdots & \vdots & \ddots
& \vdots \\
\lambda_{R, N_R}^{\frac{1}{2}}\lambda_{T,1}^{\frac{1}{2}} & \lambda_{R,N_R}^{\frac{1}{2}}\lambda_{T,2}^{\frac{1}{2}} & \cdots & \lambda_{R,N_R}^{\frac{1}{2}}\lambda_{T,N_T}^{\frac{1}{2}}
\end{bmatrix},~ \forall \mathrm{J} \in \{\mathrm{C}, \mathrm{D}\},\label{app-eq:eigenmode_coupling_matrix}
\end{equation}
\end{tiny}
\hspace{-0.15cm}where $\lambda_{R,k} \equiv 1$, for all $k \in \{1, 2,\ldots,N_R\}$, if $\mathrm{J}=\mathrm{D}$. Then, substituting \eqref{app-eq:eigenmode_coupling_matrix} into the expression of $\mathbf{M}_\mathrm{C}$ given by \eqref{eq:capacity_definition_C-MIMO-1} if $\mathrm{J}=\mathrm{C}$, or \eqref{app-eq:eigenmode_coupling_matrix} into the expression of $\mathbf{M}_\mathrm{D}$ shown in \eqref{eq:capacity_definition_D-MIMO-1} if $\mathrm{J}=\mathrm{D}$, and after performing some algebraic manipulations, we attain
\eqref{app-eq:explicit_expression_of_M} shown at the top of this page, where $\widehat{H}_{i, j}$ denotes the $(i, j)^{\mathrm{th}}$ entry of $\mathbf{\widehat{H}}$ shown in \eqref{eq:Kronecker_model}, and where $d_k=d$ and ${\varphi}_k={\varphi}$ if $\mathrm{J}=\mathrm{C}$ whereas $\lambda_{R,k}=1$ if $\mathrm{J}=\mathrm{D}$, for all $k \in \{1, 2,\ldots,N_R\}$.

In view of \eqref{app-eq:explicit_expression_of_M}, it is clear that the $(i, j)^{\mathrm{th}}$ entry, for all $i, j \in \{1, 2,\ldots,N_T\}$, of $\mathbf{M}_\mathrm{C}$ and $\mathbf{M}_\mathrm{D}$, are given by
\setcounter{equation}{40}
\begin{equation}
\label{app-eq:components_of_M}
[\mathbf{M}_\mathrm{J}]_{i, j}=\sum\limits_{k=1}^{N_R} \lambda_{T, i}^\frac{1}{2}\lambda_{T, j}^\frac{1}{2}\lambda_{R, k}d_k^{-\nu}{\varphi}_k{\widehat{H}_{k,i}}^{\rm H}\widehat{H}_{k,j},~\forall \mathrm{J} \in \{\mathrm{C}, \mathrm{D}\},
\end{equation}
where $d_k \equiv d$ and ${\varphi}_k \equiv {\varphi}$ if $\mathrm{J}=\mathrm{C}$ whereas $\lambda_{R,k} = 1$ if $\mathrm{J}=\mathrm{D}$, for all $k \in \{1, 2,\ldots,N_R\}$.

In order to apply Lemma~\ref{lemma:long-vector}, we need check the condition shown in \eqref{app_eq:long-vector-1_condition}. Specifically, for all $i, j \in \{1, 2,\ldots,N_T\}$, let
\begin{equation}
\label{app-eq:vector_gi}
\mathbf{g}_{i} \triangleq \left[a_{1,i}^\frac{1}{2}{g}_{1,i}, a_{2,i}^\frac{1}{2}{g}_{2,i},\cdots, a_{N_R,i}^\frac{1}{2}{g}_{N_R,i}\right]^{\rm T}
\end{equation}
and
\begin{equation}
\label{app-eq:vector_hj}
\mathbf{h}_{j}\triangleq \left[b_{1,j}^\frac{1}{2}{h}_{1,j}, b_{2,j}^\frac{1}{2}{h}_{2,j},\cdots, b_{N_R,j}^\frac{1}{2}{h}_{N_R,j}\right]^{\rm T},
\end{equation}
where in the \ac{C-MIMO} setting, $a_{k,i} \triangleq \lambda_{T, i}\lambda_{R, k}d^{-\nu}$, $b_{k,j} \triangleq \lambda_{T, j}\lambda_{R, k}d^{-\nu}$, ${g}_{k,i} \triangleq {\varphi}^{\frac{1}{2}}\widehat{H}_{k,i}$ and ${h}_{k,j} \triangleq {\varphi}^{\frac{1}{2}}\widehat{H}_{k,j}$, for all $k \in \{1, 2,\ldots,N_R\}$, whereas in the \ac{D-MIMO} setting, $a_{k,i} \triangleq \lambda_{T, i}d_k^{-\nu}$, $b_{k,j} \triangleq \lambda_{T, j}d_k^{-\nu}$, ${g}_{k,i} \triangleq {\varphi}_k^{\frac{1}{2}}{\widehat{H}_{k,i}}$ and ${h}_{k,j} \triangleq {\varphi}_k^{\frac{1}{2}}\widehat{H}_{k,j}$, for all $k \in \{1, 2,\ldots,N_R\}$. With these definitions in mind, it is clear that $[\mathbf{M}_\mathrm{C}]_{i, j}$ shown in \eqref{app-eq:components_of_M} equals $\mathbf{g}_{i}^{\rm H}\mathbf{h}_{j}$ where $\mathbf{g}_{i}$ and $\mathbf{h}_{j}$ are expressed as in the \ac{C-MIMO} setting, and that $[\mathbf{M}_\mathrm{D}]_{i, j}$ shown in \eqref{app-eq:components_of_M} equals $\mathbf{g}_{i}^{\rm H}\mathbf{h}_{j}$ where $\mathbf{g}_{i}$ and $\mathbf{h}_{j}$ are expressed as in the \ac{D-MIMO} setting. The correlation matrices $\mathbf{\Theta}_{T}$ and $\mathbf{\Theta}_{R}$ shown in \eqref{eq:corr_matrix} are of Toeplitz form \cite{Gray2006}. Also, it is well-known that for large numbers of antennas, the eigenvalues of $\mathbf{\Theta}$ ($\mathbf{\Theta}_{T}$ or $\mathbf{\Theta}_{R}$) converge uniformly to \cite[p. 38]{Gray2006}
\begin{equation}
\label{eq:eigenvalues_Toeplitz_large_antennas-3}
\lambda_{\mathbf{\Theta}}(x)
= (1-\theta^2)(1-2 \, \theta \cos(2\pi x)+\theta^2)^{-1},
\end{equation}
where $\theta={\theta}_{T}$ (resp. $\theta={\theta}_{R}$) for $\mathbf{\Theta}=\mathbf{\Theta}_{T}$ (resp. $\mathbf{\Theta}=\mathbf{\Theta}_{R}$), with $\theta_{T}$ and $\theta_{R}$ defined immediately after \eqref{eq:corr_matrix}, and where $x \in \left[0, 1\right]$. According to \eqref{eq:eigenvalues_Toeplitz_large_antennas-3}, the maximum eigenvalue of $\mathbf{\Theta}$ in the massive \ac{MIMO} context is then obtained by setting $x=0$ in the denominator of \eqref{eq:eigenvalues_Toeplitz_large_antennas-3}, yielding $\lambda_{\mathbf{\Theta}, \, \mathrm{max}}= (1+\theta)(1-\theta)^{-1}$, which is a finite real number, and, therefore, it is clear that there exists $b < \infty$ such that $a_{k,i}^{2}\leq b$ and $b_{k,j}^{2}\leq b$, for all $i,j \in \{1, 2,\ldots,N_T\}$ and, hence, the condition required by Lemma~\ref{lemma:long-vector} is satisfied.

Now, recalling that for all $i, j \in \{1, 2,\ldots,N_T\}$ and for all $k \in \{1, 2,\ldots,N_R\}$, $\widehat{H}_{k,i}, \widehat{H}_{k,j}\sim \mathcal{CN}(0,1)$ according to \eqref{eq:entries_of_small-scale_matrix}, it is clear that $\mathbb{E}\{[\mathbf{M}_\mathrm{C}]_{i, j}\}=\mathbb{E}\{[\mathbf{M}_\mathrm{D}]_{i, j}\}=0$, for all $i \neq j$. On the other hand, if $i=j$, by recalling the Gamma distribution of ${\varphi}$ and ${\varphi}_k$, $ k \in \{1, 2,\ldots,N_R\}$, as shown respectively in \eqref{eq:gamma_pdf_shadowing_C_MIMO} and \eqref{eq:gamma_pdf_shadowing_D_MIMO}, and recalling that  $\widehat{H}_{i, j} \sim \mathcal{CN}(0,1)$ as shown in \eqref{eq:entries_of_small-scale_matrix}, and with further algebraic manipulations, the mean of the diagonal entries of $\mathbf{M}_\mathrm{C}$ and $\mathbf{M}_\mathrm{D}$ are obtained as $\mathbb{E}\{[\mathbf{M}_\mathrm{C}]_{i, i}\}=N_R\lambda_{T, i}\, d^{-\nu}\Omega$ and $\mathbb{E}\{[\mathbf{M}_\mathrm{D}]_{i, i}\}=\lambda_{T, i}\sum_{k=1}^{N_R}d_k^{-\nu}\Omega$, respectively.

Finally, by recalling that the entries of $\mathbf{g}_{i}$ and $\mathbf{h}_{j}$ shown respectively in \eqref{app-eq:vector_gi} and \eqref{app-eq:vector_hj} depend upon the scheme (\ac{C-MIMO} or \ac{D-MIMO}) as presented right after \eqref{app-eq:vector_hj}, and noticing that $\mathbf{g}_{i}=\mathbf{h}_{j}$ if $i=j$ within each of the schemes, and assuming $N_R \to \infty$, we apply Lemma~\ref{lemma:long-vector} to the vectors $\mathbf{g}_{i}$ and $\mathbf{h}_{j}$, yielding \eqref{app-eq:asymp_behav_of_non_diag_entries_of_Mc} and \eqref{app-eq:asymp_behav_of_diag_entries_of_Mc} shown at the top of this page, where $d_k \equiv d$, for all $k \in \{1, 2,\ldots,N_R\}$, if $\mathrm{J}=\mathrm{C}$. By virtue of \eqref{app-eq:asymp_behav_of_non_diag_entries_of_Mc} and \eqref{app-eq:asymp_behav_of_diag_entries_of_Mc}, we attain \eqref{app-eq:asymp_behav_diagonal_of_Md}, shown also at the top of this page, where $d_k \equiv d$, for all $k \in \{1, 2,\ldots,N_R\}$, if $\mathrm{J}=\mathrm{C}$. This completes the proof.

\section{Proof of Eq. \eqref{eq:ergodic_capacity_D-MIMO_circular}}
\label{Proof_of_equation_ergodic_capacity_D-MIMO_circular}
In view of Fig. \ref{fig:system_model_circular}-b and by recalling the well-known law of cosines in geometry, we have, for all $k \in \{1, 2,\ldots,N_R\}$, $d_k=\left[r_{\mathrm{u}}^{2}+r_{\mathrm{a}}^{2}-2r_{\mathrm{u}}r_{\mathrm{a}} \cos(\phi_k-\phi)\right]^{1/2}=\left(r_{\mathrm{u}}^{2}+r_{\mathrm{a}}^{2}-2r_{\mathrm{u}}r_{\mathrm{a}} \cos\omega_k\right)^{1/2}$, where $\omega_k\triangleq\phi_k-\phi$ is the angle between the segments $OU$ and $OL_k$. Let $\Delta\omega\triangleq\omega_{k+1}-\omega_k$. Then, by recalling that $\phi_1=0$ and that $\phi_k=2\pi(k-1)/{N_R}$, for all $k \in [2, N_R]$ (cf. Section \ref{subsec:system_circular}), we have $\Delta\omega = 2\pi/{N_R}$. Therefore, if $N_R\to\infty$, we have
\setcounter{equation}{47}
\begin{align}
\frac{1}{N_R}\sum\limits_{k=1}^{N_R}d_k^{-\nu}
&=\frac{1}{N_R \, \Delta\omega}\sum\limits_{k=1}^{N_R}\left(r_{\mathrm{u}}^{2}+r_{\mathrm{a}}^{2}-2r_{\mathrm{u}}r_{\mathrm{a}} \cos\omega_k\right)^{-\nu/2}\Delta\omega \label{eq:sigma_dk_circular-1} \\
& \xrightarrow{N_R \to \infty} \frac{1}{2\pi}\int_{0}^{2\pi}\left(r_{\mathrm{u}}^{2}+r_{\mathrm{a}}^{2}-2r_{\mathrm{u}}r_{\mathrm{a}} \cos\omega\right)^{-\nu/2} \mathrm{d}\omega \label{eq:sigma_dk_circular-3}\\
&=\frac{1}{2\pi}\int_{-\pi}^{\pi}\left(r_{\mathrm{u}}^{2}+r_{\mathrm{a}}^{2}-2r_{\mathrm{u}}r_{\mathrm{a}} \cos\omega\right)^{-\nu/2}\mathrm{d}\omega \label{eq:sigma_dk_circular-4},
\end{align}
where the identity $\int_{0}^{2\pi} \cos\omega \, \mathrm{d}\omega=\int_{-\pi}^{\pi}\cos\omega \, \mathrm{d}\omega$ was exploited to derive \eqref{eq:sigma_dk_circular-4}. Finally, applying \cite[Eq. (2.5.16.38)]{Prudnikov86_elementary} to \eqref{eq:sigma_dk_circular-4} and performing some algebraic manipulations yields \eqref{eq:ergodic_capacity_D-MIMO_circular}.

\section{Proof of Theorem \ref{theorem:average_asymptotic_behavior_circular_C-MIMO_and_D-MIMO}}
\label{Proof_of_average_asymptotic_behavior_circular_C-MIMO_and_D-MIMO}

In the \ac{D-MIMO} scheme, to obtain simple yet meaningful expression of the average asymptotic spectral efficiency of the system, we consider the range of medium-to-high \ac{SNR}, where $\beta_i=\rho\Omega N_R\lambda_{T,i}N_{T}^{-1} \overset{N_R \to \infty} \longrightarrow \infty$, for all $i \in  \{1, 2,\ldots,N_T\}$. The average asymptotic spectral efficiency of the system in this case is then derived as
\begin{align}
\overline{C}_\mathrm{D, \, circ}^{\mathrm{avg}}
&= \int_{0}^{r_{\mathrm{c}}}f_{r_{\mathrm{u}}}(x)\overline{C}_\mathrm{D, \, circ}(r_{\mathrm{u}}) \, \mathrm{d}{r_{\mathrm{u}}} \label{app-eq:average_asymptotic_behavior_circular_D-MIMO-1}\\
 &\xrightarrow{p.} \frac{2}{r_{\mathrm{c}}^2\ln2}\sum\limits_{i=1}^{N_T}\int_{0}^{r_{\mathrm{c}}}r_{\mathrm{u}}\ln\left[1+\beta_i\frac{r_{\mathrm{u}}^{2}+r_{\mathrm{a}}^{2}}{|r_{\mathrm{u}}^{2}-r_{\mathrm{a}}^{2}|^{3}}\right] \, \mathrm{d}{r_{\mathrm{u}}} \label{app-eq:average_asymptotic_behavior_circular_D-MIMO-2} \\
&\approx \frac{1}{r_{\mathrm{c}}^2\ln2}\sum\limits_{i=1}^{N_T}\int_{0}^{r_{\mathrm{c}}^{2}} \ln\left[\beta_i\frac{y+r_{\mathrm{a}}^{2}}{|y-r_{\mathrm{a}}^{2}|^{3}}\right] \, \mathrm{d}y \label{app-eq:average_asymptotic_behavior_circular_D-MIMO-3}\\
&= \frac{1}{r_{\mathrm{c}}^2\ln2}\!\sum\limits_{i=1}^{N_T}\!\left\{\! \int_{0}^{r_{\mathrm{c}}^{2}}\!\!\!\ln\beta_i  \, \mathrm{d}y \!+\!\!\int_{0}^{r_{\mathrm{c}}^{2}}\!\!\!\ln(y+r_a^{2}) \, \mathrm{d}y\right. \nonumber \\
&    \quad  \left. -3 \!\!\int_{0}^{r_{\mathrm{a}}^{2}}\!\!\!\ln(r_a^{2}-y) \, \mathrm{d}y
-3\!\!\int_{r_{\mathrm{a}}^{2}}^{r_{\mathrm{c}}^{2}}\!\!\!\!\ln(y-r_a^{2}) \, \mathrm{d}y \right\}
\label{app-eq:average_asymptotic_behavior_circular_D-MIMO-4},
\end{align}
where the approximation $\ln(1+x) \approx \ln x$ if $x \to \infty$ was used along with the change of variable $y=r_{\mathrm{a}}^{2}$, to obtain \eqref{app-eq:average_asymptotic_behavior_circular_D-MIMO-3}. Finally, solving the integrals involved in \eqref{app-eq:average_asymptotic_behavior_circular_D-MIMO-4} and performing further algebraic manipulations, leads to the intended \eqref{eq:average_asymptotic_behavior_circular_D-MIMO}.

\section{Proof of Corollary \ref{corollary:optimal_location_antenna_array_D-MIMO}}
\label{Proof_of_corollary:optimal_location_antenna_array_D-MIMO}

The first-order derivative of the average asymptotic spectral efficiency, i.e., $\overline{C}_\mathrm{D, \, circ}^{\mathrm{avg}}$ given by \eqref{eq:average_asymptotic_behavior_circular_D-MIMO}, w.r.t. the antenna array size $r_{\mathrm{a}}$, is given by
\begin{equation}
\label{app-eq:first_derivative_average_asymptotic_behavior_circular_D-MIMO}
\frac{\mathrm{d}}{\mathrm{d}r_{\mathrm{a}}}\overline{C}_\mathrm{D, \, circ}^{\mathrm{avg}}=6N_T\frac{r_{\mathrm{a}}}{r_{\mathrm{c}}^2}\log_2(\chi-2)+2N_T\frac{r_{\mathrm{a}}}{r_{\mathrm{c}}^2}\log_2\chi,
\end{equation}
where $\chi \triangleq r_{\mathrm{c}}^2/r_{\mathrm{a}}^2 + 1$.
Then, by setting $\frac{\mathrm{d}}{\mathrm{d}r_{\mathrm{a}}}\overline{C}_\mathrm{D, \, circ}^{\mathrm{avg}}=0$ and performing some algebraic manipulations, we get the equation $(r_{\mathrm{c}}^2-r_{\mathrm{a}}^2)^{3}(r_{\mathrm{c}}^2+r_{\mathrm{a}}^2)-r_{\mathrm{a}}^8=0$, which can be reformulated as
\begin{eqnarray}
\chi(\chi-2)^{3}-1=0 \, \, \iff \, \, \chi^{\frac{4}{3}}-2\chi^{\frac{1}{3}}-1=0 \label{eq:equiv_equation_r_optimal-2}.
\end{eqnarray}
After performing some algebraic manipulations, the solution to \eqref{eq:equiv_equation_r_optimal-2}, denoted $\chi_0$, can be readily shown as $\chi_0=2.7167$. Finally, by using the definition of $\chi$ right after \eqref{app-eq:first_derivative_average_asymptotic_behavior_circular_D-MIMO}, we attain \eqref{eq:optimal_antenna_array_dimension}.

On the other hand, by multiplying both sides of \eqref{app-eq:first_derivative_average_asymptotic_behavior_circular_D-MIMO} by $1/r_{\mathrm{a}}$, it is clear that the function $\frac{1}{r_{\mathrm{a}}}\frac{\mathrm{d}}{\mathrm{d}r_{\mathrm{a}}}\overline{C}_\mathrm{D, \, circ}^{\mathrm{avg}}$ is decreasing with $r_{\mathrm{a}}$, then $\frac{1}{r_{\mathrm{a}}}\frac{\mathrm{d}}{\mathrm{d}r_{\mathrm{a}}}\overline{C}_\mathrm{D, \, circ}^{\mathrm{avg}}<0$ if $r_{\mathrm{a}}<r_{\mathrm{a}}^{\mathrm{opt}}$, and $\frac{1}{r_{\mathrm{a}}}\frac{\mathrm{d}}{\mathrm{d}r_{\mathrm{a}}}\overline{C}_\mathrm{D, \, circ}^{\mathrm{avg}}>0$ if $r_{\mathrm{a}}>r_{\mathrm{a}}^{\mathrm{opt}}$. Since $r_{\mathrm{a}} > 0$, we have $\frac{\mathrm{d}}{\mathrm{d}r_{\mathrm{a}}}\overline{C}_\mathrm{D, \, circ}^{\mathrm{avg}}<0$ if $r_{\mathrm{a}}<r_{\mathrm{a}}^{\mathrm{opt}}$, and $\frac{\mathrm{d}}{\mathrm{d}r_{\mathrm{a}}}\overline{C}_\mathrm{D, \, circ}^{\mathrm{avg}}>0$ if $r_{\mathrm{a}}>r_{\mathrm{a}}^{\mathrm{opt}}$, which concludes that $r_{\mathrm{a}}^{\mathrm{opt}}$ is the maximum of the average asymptotic spectral efficiency.

\bibliographystyle{IEEEtran}
\bibliography{IEEEabrv,./References}

\begin{IEEEbiography}
[{\includegraphics[width=1in, height=1.25in,  clip, keepaspectratio]{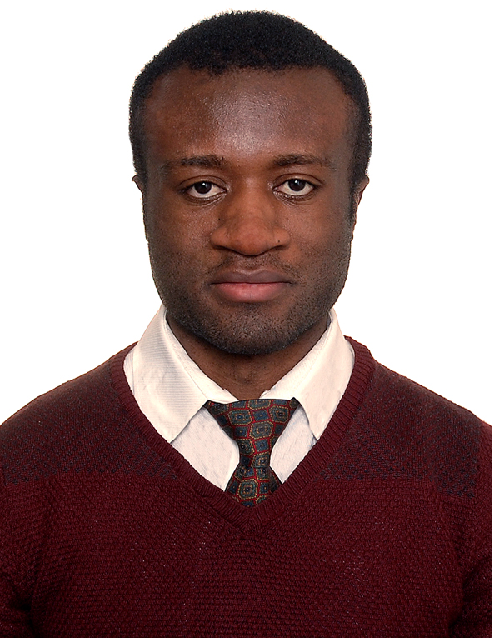}}]{Gervais N. Kamga} (S'13) received the B.S. degree in Electrical and Computer Engineering from Ecole Mohammadia d'Ing\'enieurs (EMI), Rabat, Morocco, in 2011. He joined the Institut National de la Recherche Scientifique (INRS), University of Quebec, Montreal, QC, Canada, where he is currently working towards the Ph.D. degree in Telecommunications.

Gervais was a prize winner of the Pan African Mathematics Olympiads in 2006. He was a recipient of the Excellence Scholarship for International Cooperation Morocco-Cameroon, and the Financial Assistance of the State of Cameroon to Scholarship Students Abroad, from 2006 to 2011. Between 2010 and 2011, he won, respectively, the National Students Open Design and Innovation Competition (SODEC), and the National Competition for Technological Innovation (EMINOV), Morocco. He was also a finalist in the 2010 EMINOV. In 2011, he was a recipient of the International ParisTech and Renault Foundation scholarship for the Mobility and Electric Vehicles MasterÕs Program. At INRS, he is recipient of a waiver of high tuition fees for foreign students, and university scholarship for graduate studies. He was, in 2013, a semi-finalist to the Quebec National Concours Forces Avenir, in the Sciences and Technology Applications category.
\end{IEEEbiography}

\begin{IEEEbiography}
[{\includegraphics[width=1in, height=1.25in,  clip, keepaspectratio]{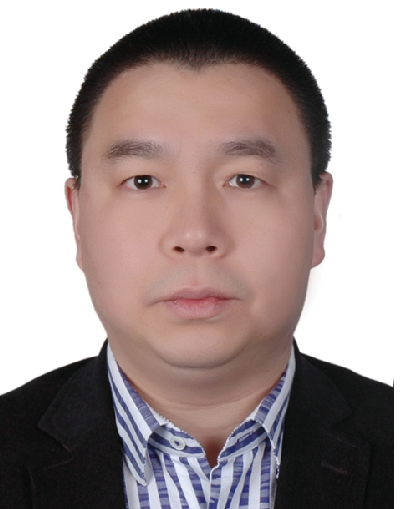}}]{Minghua Xia} (M'12) obtained his Ph.D. degree in Telecommunications and Information Systems from Sun Yat-sen University, Guangzhou, China, in 2007. Since 2015, he has been working as a Professor at the same university.

From 2007 to 2009, he was with the Electronics and Telecommunications Research Institute (ETRI) of South Korea, Beijing R\&D Center, Beijing, China, where he worked as a member and then as a senior member of engineering staff and participated in the projects on the physical layer design of 3GPP LTE mobile communications. From 2010 to 2014, he was in sequence with The University of Hong Kong, Hong Kong, China; King Abdullah University of Science and Technology, Jeddah, Saudi Arabia; and the Institut National de la Recherche Scientifique (INRS), University of Quebec, Montreal, Canada, as a Postdoctoral Fellow. His research interests are in the general area of 5G wireless communications, and in particular the design and performance analysis of multi-antenna systems, cooperative relaying systems and cognitive relaying networks, and recently focus on the design and analysis of wireless power transfer and/or energy harvesting systems, as well as massive MIMO and small cells. He holds two patents granted in China.

Dr. Xia received the Professional Award at IEEE TENCON'15, Macau, 2015. He was also awarded as an Exemplary Reviewer by {\scshape IEEE Transactions on Communications}, {\scshape IEEE Communications Letters}, and {\scshape IEEE Wireless Communications Letters}, respectively, in 2014.
\end{IEEEbiography}

\begin{IEEEbiography}
[{\includegraphics[width=1in, height=1.25in,  clip, keepaspectratio]{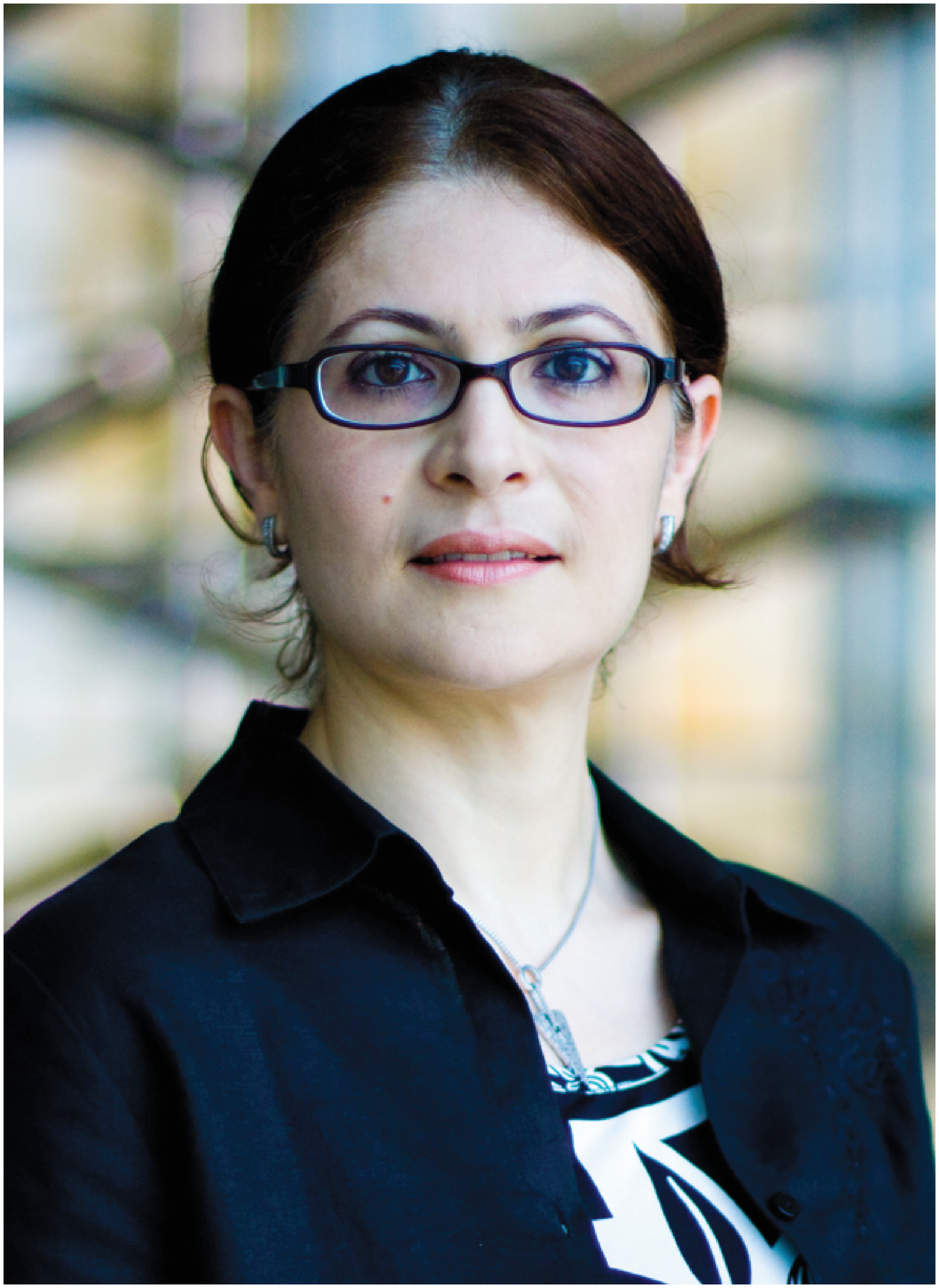}}]{Sonia A\"{\i}ssa} (S'93-M'00-SM'03) received her Ph.D. degree in Electrical and Computer Engineering from McGill University, Montreal, QC, Canada, in 1998. Since then, she has been with the Institut National de la Recherche Scientifique-{\it Energy, Materials and Telecommunications} Center (INRS-EMT), University of Quebec, Montreal, QC, Canada, where she is a Full Professor.

From 1996 to 1997, she was a Researcher with the Department of Electronics and Communications of Kyoto University,
and with the Wireless Systems Laboratories of NTT, Japan. From 1998 to 2000, she was a Research Associate at INRS-EMT. In 2000-2002, while she was an Assistant Professor, she was a Principal Investigator in the major program of personal and mobile communications of the Canadian Institute for Telecommunications Research, leading research in radio resource management for wireless networks. From 2004 to 2007, she was an Adjunct Professor with Concordia University, Montreal. She was Visiting Invited Professor at Kyoto University, Japan, in 2006, and Universiti Sains Malaysia, in 2015.
Her research interests include the modeling, design and performance analysis of wireless communication systems and networks.

Dr. A\"{\i}ssa is the Founding Chair of the IEEE Women in Engineering Affinity Group in Montreal, 2004-2007; acted as TPC Symposium Chair or Cochair at IEEE ICC '06 '09 '11 '12; Program Cochair at IEEE WCNC 2007; TPC Cochair of IEEE VTC-spring 2013; and TPC Symposia Chair of IEEE Globecom 2014. Her main editorial activities include: Editor, {\scshape IEEE Transactions on Wireless Communications}, 2004-2012; Associate Editor and Technical Editor, {\scshape IEEE Communications Magazine}, 2004-2015; Technical Editor, {\scshape IEEE Wireless Communications Magazine}, 2006-2010; and Associate Editor, {\it Wiley Security and Communication Networks Journal}, 2007-2012. She currently serves as Area Editor for the {\scshape IEEE Transactions on Wireless Communications}. Awards to her credit include the NSERC University Faculty Award in 1999; the Quebec Government FRQNT Strategic Faculty Fellowship in 2001-2006; the INRS-EMT Performance Award multiple times since 2004, for outstanding achievements in research, teaching and service; and the Technical Community Service Award from the FQRNT Centre for Advanced Systems and Technologies in Communications, 2007. She is co-recipient of five IEEE Best Paper Awards and of the 2012 IEICE Best Paper Award; and recipient of NSERC Discovery Accelerator Supplement Award. She is a Distinguished Lecturer of the IEEE Communications Society (ComSoc) and an Elected Member of the ComSoc Board of Governors. Professor A\"{\i}ssa is a Fellow of the Canadian Academy of Engineering.
\end{IEEEbiography}

\vfill

\end{document}